
\input phyzzx
\PHYSREV

\def\wdhat{\mathaccent"0362}
\def\wdtil{\mathaccent"0365}

\def\ftil{{\wdtil{F}}}
\def\rtil{{\wdtil{R}}}\def\rt{{\wdtil{r}}}
\def\br{\overline}

\def\lag{{\cal L}}

\def\detg{\sqrt{-\rm{g}}}

\def\a2{\left| A\right|^{2} }
\def\f2{\left| F\right|^{2} }
\def\h2{\left| H\right|^{2} }

\def\eabc{\epsilon_{abc}}
\def\fmni{ F^{(i)}_{\mu\nu}}
\def\fimn{ F^{(i)\mu\nu}}

\def\rr{R\left( r\right)}
\def\tr{T\left( r\right)}
\def\mr{m\left( r\right)}
\def\R2{\left(1-{2G\mr\over r} \right)}
\def\drr{-\ln\left( R/ T\right)}
\def\dr{\delta\left( r\right)}
\def\edr{e^{-2\delta}}
\def\brr{\left( r\right)}
\def\brh{\left( r_{h}\right)}
\def\br0{\left( 0\right)}
\def\brinf{\left( \infty\right)}
\def\dom{\left(d\theta^{2} +\sin^{2}\theta d\varphi^2\right)}
\def\dx{d^4 x}
\def\pd{\partial}
\def\b2{b_{\mu}b^{\mu}}
\def\tar{{\wdhat \tau}_{r}}
\def\tat{{\wdhat \tau}_{\theta}}
\def\tap{{\wdhat \tau}_{\varphi}}
\def\taa{{\wdhat \tau}_{a}}
\def\tab{{\wdhat \tau}_{b}}
\def\tac{{\wdhat \tau}_{c}}

\def\c1{\left( 1+c\right)}
\def\w1{\left( 1+f\right)}
\def\ww1{\left( 1-f^2\right)}
\def\brt{\beta\left( r, t\right)}

\def\wp2{\left( f^{\prime}\right)^2}

\def\drh{{\wdhat {dr}}}
\def\dth{{\wdhat {d\theta}}}
\def\dph{{\wdhat {d\varphi}}}
\def\gint{\int \dx\detg}

\def\rhat{\wdhat r}

\def\mhat{\wdhat m}
\def\rrh{\left( r-r_{h}\right)}
\def\wwrh1{\left[ 1-f^2\brh\right]}
\def\wrh1{\left[ 1+f\brh\right]}

\def\ep{e^{-2\phi}}
\def\p0{\phi_{\infty}}
\def\eyr{e^{-2\left(  h/r +\p0\right)}}
\def\bryr{\left( h/ r\right)}

\def\yr{{h\over r}}

\magnification=1000
\FRONTPAGE
{\singlespace\centerline{
{}~~\hfil~~~~\hfil~~~~\hfil~~~~\hfil~~~~\hfil~~~~\hfil~~
{}~~\hfil~~~~\hfil~~~~\hfil~~~~\hfil~~~~\hfil~~~~\hfil~~
CLNS-93/1246~}}
{\singlespace\centerline{
{}~~\hfil~~~~\hfil~~~~\hfil~~~~\hfil~~~~\hfil~~~~\hfil~~
{}~~\hfil~~~~\hfil~~~~\hfil~~~~\hfil~~~~\hfil~~~~\hfil~~
hep-th/9311022 ~}}
{\singlespace\centerline{
{}~~\hfil~~~~\hfil~~~~\hfil~~~~\hfil~~~~\hfil~~~~\hfil~~
{}~~\hfil~~~~\hfil~~~~\hfil~~~~\hfil~~~~\hfil~~~~\hfil~~
November 1993}}
\vskip 1.1in
{\singlespace
\centerline{\fourteenpoint{\bf Einstein-Yang-Mills Theory with
 a}}
\centerline{\fourteenpoint{\bf Massive Dilaton and Axion:
String-Inspired}}
\centerline{\fourteenpoint {\bf Regular and Black Hole Solutions}}
\centerline{ }

{\singlespace\centerline{ }
\centerline{\twelvepoint   Christopher M. O'Neill}
\centerline{ }
 \centerline{\twelvepoint Laboratory of Nuclear Studies}
\centerline{\twelvepoint Cornell University}
\centerline{\twelvepoint Ithaca, NY  14853 }
\centerline{ }
\normalspace
\centerline{ }
\centerline{ }
\centerline{ }
\centerline{\fourteenpoint Abstract}

\REF\bek{J.D.~Bekenstein
\journal Phys.Rev.&D~5(1972)1239.}
\REF\adpe{S.L.~Adler and R.D.~Pearson
\journal Phys.Rev.&D~18(1978)2798}
\REF\teit{C.~Teitelboim
\journal Phys.Rev.&D~5(1972)2941.}
\REF\price{R.H.~Price
\journal Phys.Rev.&D~5(1972)2419.}
\REF\deser{S.~Deser
\journal Class.Quantum Grav.&1(1984)L1.}
\REF\colea{S.~Coleman
\journal Commun.Math.Phys.&55(1977)113.}
\REF\biz{P.~Bizon
\journal Phys.Rev.Lett.&64(1990)2844.}
\REF\voga{M.S.~Volkov and D.V.~Gal'tsov
\journal Yad.Fiz.&51(1990)1171
\journal [Sov.J.Nucl.Phys.&51(1990)747].}
\REF\kuma{H.P.~Kunzle and A.K.M.~Masood-ul-Alam
\journal J.Math.Phys.&31(1990)928.}
\REF\bart{R.~Bartnik and J.~McKinnon
\journal Phys.Rev.Lett.&61(1988)141.}
\REF\erga{A.A.~Ershov and D.V.~Gal'tsov
\journal Phys.Lett.&A~150(1990)159.}
\REF\galvol{D.V.~Gal'tsov and M.S.~Volkov
\journal Phys.Lett.&A~162(1992)144.}
\REF\suwa{D.~Sudarsky and R.~Wald,
\journal Phys.Rev.&D~46(1992)1453.}
\REF\stzha{N.~Straumann and Z.~Zhou
\journal Phys.Lett.&B~237(1990)35.}
\REF\stzhb{N.~Straumann and Z.~Zhou
\journal Phys.Lett.&B~243(1990)33.}
\REF\bizoa{P.~Bizon
\journal Phys.Lett.&B~259(1991)53.}
\REF\stzhc{Z.~Zhou and N.~Straumann
\journal Nucl.Phys.&B~360(1991)180.}
\REF\coprwia{S.~Coleman, J.~Preskill, and F.~Wilczek
\journal Nucl.Phys.&B~378(1992)175.}
\REF\coprwib{ S.~Coleman, J.~Preskill and F.~Wilczek
\journal Phys.Rev.Lett&67(1991)1975.}
\REF\lenawe{K.~Lee, V.P.~Nair, and E.J.~Weinberg
\journal Phys.Rev.&D~45(1992)2751.}
\REF\ortiz{M.~Ortiz\journal Phys.Rev.&D~45(1992)2586.}
\REF\brfoma{P.~Breitenlohner, P.~Forg\' acs and D.~Maison
\journal Nucl.Phys.&B~383(1992)357.}
\REF\grmaon{B.R.~Greene, S.D.~Mathur and C.M.~O'Neill
\journal Phys.Rev.&D~47(1993)2242.}
\REF\dahane{R.F.~Dashen, B.~Hasslacher and A.~Neveu
\journal Phys.Rev.&D~10(1974)4138.}
\REF\klma{F.R.~Klinkhamer and N.S.~Manton
\journal Phys.Rev.&D~30(1984)2212.}
\REF\hedrsta{M.~Heusler, S.~Droz,and N.~Straumann
\journal Phys.Lett.&B~285(1992)21.}
\REF\hedrstb{S.~Droz, M.~Heusler, and N.~Straumann
\journal Phys.Lett.&B~268(1991)371.}
\REF\bich{P.~Bizon and T.Chmaj
\journal Phys.Lett.&B~297(1992)55.}
\REF\katr{ D.~Kastor and J.~Traschen
\journal Phys.Rev.&D~46(1992)5399.}
\REF\gima{ G.~Gibbons and K.~Maeda
\journal Nucl.Phys.&B~298(1988)741.}
\REF\gahost{D.~Garfinkle, G.~Horowitz and A.~Strominger
\journal Phys.Rev.&D~43(1991)3140.}
\REF\prscshtrwi{J.~Preskill,  P.~Schwarz, A.~Shapere, S.~Trivedi
and F.~Wilczek
\journal Mod.Phys.Lett.&A~6(1991) \overfullrule0pt 2353.}
\REF\shtrwi{A.~Shapere, S.~Trivedi and F.~Wilczek
\journal Mod.Phys.Lett.&A~6(1991)2677.}
\REF\howi{C.~Holzey and F.~Wilczek
\journal Nucl.Phys&B~380(1992)447.}
\REF\hohoa{J.H.~Horne and G.T.~Horowitz
\journal Phys.Rev.&D~46(1992)1340.}
\REF\cakaola{B.A.~Campbell, N.~Kaloper and K.A.~Olive
\journal Phys.Lett&B~263(1991)364.}
\REF\lewe{ K.~Lee and E.J.~Weinberg
\journal Phys.Rev.&D~44(1991)3159.}
\REF\cdkoa{B.A.~Campbell, M.J.~Duncan, N.~Kaloper and K.A.~Olive
\journal Phys.Lett.&B~251(1990)148.}
\REF\mist{S.~Mignemi and N.R.~Stewart
\journal Phys.Lett.&B~298(1993)299.}
\REF\asena{A.~Sen
\journal Phys.Rev.Lett.&69(1992)1006.}
\REF\cakaolb{B.A.~Campbell, N.~Kaloper and K.A.~Olive
\journal Phys.Lett.&B~285(1992)199.}
\REF\ghor{ G.~Horowitz,
``The Dark Side of String Theory: Black Holes and Black Strings'',
 preprint UCSBTH-92-32, October 1992.}
\REF\asen{A.~Sen,
``Black Holes and Solitons in String Theory'',
preprint TIFR-TH-92-57, October 1992.}
\REF\hohob{J.H.~Horne and G.T.~Horowitz
\journal Nucl.Phys.&B~399(1993)169.}
\REF\grha{R.~Gregory and J.A.~Harvey
\journal Phys.Rev.&D~47(1993)2411.}
\REF\lamaa{G.~Lavrelashvili and D.~Maison,
``Regular and Black Hole Solutions of Einstein--Yang-Mills--Dilaton
Theory'', preprint MPI-PH-92-115, December 1992.}
\REF\lamab{G.~Lavrelashvili and D.~Maison,
``Dilatonic Sphalerons and Non-Abelian Black Holes'',
preprint MPI-PH-93-55, July 1993.}
\REF\bizob{P.~Bizon
\journal Act.Phys.Polon.&B~24(1993)1209.}
\REF\dogaa{E.E.~Donets and D.V.~Gal'tsov
\journal Phys.Lett.&B~302(1993)411.}
\REF\dogab{E.E.~Donets an D.V.~Gal'tsov,
``Charged Stringy Black Holes with Non-Abelian Hair'',
preprint DTP-MSU-93-01, April 1993.}
\REF\lamac{G.~Lavrelashvili and D.~Maison,
\journal Phys.Lett.&B~295(1992)67.}
\REF\bizoc{P.~Bizon
\journal Phys.Rev.&D~47(1993)1656.}
\REF\hest{M.~Heusler and N.~Straumann
\journal Class.Quantum Grav.&9(1992)2177.}
\REF\grsl{ D.J.~Gross and J.H.~Sloan
\journal Nucl.Phys.&B~291(1987)41.}
\REF\bodea{ D.G.~Boulware and S.~Deser
\journal Phys.Rev.Lett.&55(1985)2656.}
\REF\whe{ J.T.~Wheeler
\journal Nucl.Phys.&B~268(1986)737.}
\REF\wett{ C.~Wetterich
\journal Nucl.Phys.&B~324(1989)141.}
\REF\bodeb{ D.G.~Boulware and S.~Deser
\journal Phys.Lett.&B~175(1986)409.}
\REF\wuwa{ Y.S.~Wu and Z.~Wang
\journal Phys.Rev.Lett.&57(1986)1978.}
\REF\anco{ I.~Antoniadis and C.~Kounnas
\journal Nucl.Phys.&B~284(1987)729.}
\REF\kakool{ S.~Kalara,C.~Kounnas and K.A.~Olive
\journal Phys.Lett.&B~215(1988)265.}
\REF\kaol{S.Kalara and K.A.~Olive
\journal Phys.Lett.&B~218(1989)148.}
\REF\koyo{T.~Koikawa and M.~Yoshimura
\journal Phys.Lett.&B~189(1987)29.}
\REF\gibb{ G.~Gibbons
\journal Nucl.Phys.&B~207(1982)337.}
\REF\cdkob{B.A.~Campbell, M.J.~Duncan, N.~Kaloper and K.A.~Olive
\journal Nucl.Phys.&B~351(1991)778.}
\REF\drsw{M.~Dine, R.~Rohm, N.~Seiberg and E.~Witten
\journal Phys.Lett.&B~156(1985)55.}
\REF\mapa{K.~Maeda and P.Y.T.~Pang
\journal Phys.Lett.&B~180(1986)29.}
\REF\biga{P.~Bin\' etruy and M.K.~Gaillard
\journal Phys.Rev.&D~34(1986)3069.}
\REF\kaolb{N.Kaloper and K.A.~Olive,
\journal Astropart.Phys&1(1993)185.}
\REF\wit{E.~Witten
\journal Phys.Rev.Lett.&38(1977)121.}
\REF\cmo{C.M.~O'Neill, paper in preparation.}

We study the classical theory of a non-Abelian gauge field (gauge group
 $SU(2)$) coupled to a massive dilaton, massive axion
and Einstein gravity. The theory is inspired by the bosonic part of
 the low-energy heterotic string action for a general Yang-Mills field,
 which we consider to leading order after
compactification to $(3+1)$ dimensions. We impose the condition that
spacetime be static and spherically symmetric, and we introduce masses
via a dilaton-axion potential associated with supersymmetry
(SUSY)-breaking by gaugino condensation in the hidden sector. In the
course of describing the possible non-Abelian solutions of the
simplified theory, we consider in detail two candidates:
a massive dilaton coupled to a purely magnetic Yang-Mills field, and
a massive axion field coupled to a non-Abelian dyonic configuration,
in which the electric and magnetic fields decay too rapidly to
correspond to any global gauge charge.
We discuss the feasibility of
solutions with and without a nontrivial dilaton for the latter case,
and present numerical regular and black hole solutions for the former.

\vfill\supereject

{\parindent=0pt\bf\chapter{ Introduction{}~\hfil~~}}

Following the early investigation of a variety of field theories
coupled to Einstein gravity [\bek -\price ], it was widely believed
that only the charges carried by massless gauge fields could
characterize the exterior of a black hole. The notion that mass,
angular momentum, and ``electric'' and ``magnetic'' charges are the
only distinguishing features outside the horizon became known as the
 no-hair conjecture.

In light of the no-hair results and several no-go theorems for
classical glueball solutions with [\deser ] and without [\colea ]
gravity, the recent discovery of both black hole [\biz -\kuma ] and
smooth [\bart ] solutions of $SU(2)$ gauge theory coupled to Einstein
gravity came as quite a surprise. The fields in such solutions decay
sufficiently quickly that no global gauge charges are present, and
hence no imprint at spatial infinity is required for the existence of
 nontrivial gauge field structure. It was later shown that static
solutions {\it with} global electric or magnetic gauge charges can
only occur in the embedded Abelian sector of this theory, and that
non-Abelian dyons and dyonic black holes are prohibited [\erga ].
Further analysis has also established a sphaleron interpretation of
some smooth solutions which bridge topologically distinct Yang-Mills
vacua [\galvol -\suwa ], and the inherent instability of such
saddle-point field configurations may help explain the generic
instability of all Einstein-Yang-Mills (EYM) solutions against
collapse into Schwarzschild black holes [\stzha -\stzhc ]. Despite
their lack of stability, these non-Abelian solutions still present a
challenge to the no-hair results, which are not based on the issue of
 stability, but rather follow from the careful analysis of several
theories which are fundamentally different in character from EYM
theory.

In fact, the advent of these solutions has helped inspire a re-thinking
 of the no-hair conjecture, as well as a wealth of other solutions
incorporating non-Abelian structure. In [\coprwia ], a distinction is
drawn between primary hair, such as the structure arising from the
familiar continuous gauge charges, and secondary hair [\coprwib],
which exists solely as a result of primary hair sources and hence
does not constitute a fundamentally new characteristic\foot{ We do not
discuss quantum hair in this paper, although this distinction applies
equally well to the quantum and classical cases.}. This distinction is
well illustrated by the two recent approaches to Einstein-Yang-Mills
-Higgs (EYMH) theory: for the ``black holes inside magnetic
monopoles'' of [\lenawe -\brfoma ], the 't Hooft-Polyakov monopole
charge supports a core of secondary (triplet) Higgs hair outside the
 horizon, while for the case of $SU(2)$ coupled to a Higgs doublet
(the massive vector theory of the standard model less hypercharge)
[\grmaon ], gauge and Higgs hair exist near the horizon without global
gauge or topological charges. Thus, a distant observer in the latter
case would not be able to distinguish such an object from a
Schwarzschild black hole of the same mass, which motivates an alternate
 definition of primary hair: when the properties of a black hole are
 no longer completely determined {\it within a given theory} by the
mass, angular momentum and continuous gauge charges, the additional
parameters required to describe the black hole expand the space of
states and give rise to primary hair [\coprwia ].

Though the latter case and the original EYM black holes are examples
of such primary hair, neither seem to share the stability properties
of the well-known primary hair solutions. The black hole solutions of
 the spontaneously broken gauge theory appear to be unstable because
of their similarity to the sphaleron-like EYM solutions of [\bart ,
\biz -\kuma ], and their interpretation as gravitating generalizations
of the familiar $SU(2)$ sphalerons: the weak-gravity limit of one
class of solutions is equivalent to the YMH configuration of
[\dahane -\klma ]. On the other hand, when we ignore Hawking radiation,
 the secondary hair solutions of [\lenawe -\brfoma ] {\it are} stable
 for the same reasons flat-space monopoles are
stable. Thus, the physically important condition of stability appears
to be more closely tied to the stability properties of corresponding
flat-space solitons (when such solutions exist) than to the
classification of structure as primary or secondary hair. Other
illustrations of this point have been found (e.g.~the linearly
stable [\hedrsta ] black hole solutions to Einstein-Skyrme theory
[\hedrstb -\bich ]), and a systematic approach to the existence of
 black hole solutions with solitonic flat-space counterparts has
been formulated [\katr ], but a systematic treatment of the stability
of such solutions is currently lacking. The search continues for
stable black hole solutions to physically relevant theories, even
several years after the first challenge to the no-hair conjecture
opened the door to rich, new structure in the exterior of black holes.

On a separate front, some recent progress in black hole physics has
stemmed from the generic modifications to gravity mandated by string
theory, a promising candidate for a consistent theory of quantum
gravity which also provides predictions that challenge general
relativity well below Planck scale curvatures. In particular, the
presence at low energies of the dilaton and axion, two scalars with
 unusual couplings which appear in the same supersymmetric multiplet
as the graviton, has precipitated a host of new black hole solutions
with secondary hair and interesting properties. In charged dilaton
black holes [\gima -\hohoa ], the Maxwell field acts as a source for
 dilaton hair, which leads to modifications of causal structure that
 help shed light on several puzzles peculiar to the Reissner-Nordstrom
spacetime, as well as some mysteries of the later stages of Hawking
evaporation. Because the axion couples to $F
\ftil\sim {\bf E} \cdot {\bf B}$, black holes with both electric and
magnetic charge can support axion hair [\cakaola,\lewe ]. Another
axion coupling is of the Lorentz Chern-Simons form, so that background
metrics reflecting nonzero angular momentum can give rise to axion hair
 [\cdkoa ], which in turn acts as a source for dilaton hair [\mist ],
 without  the need for $U(1)$ charges. The more general case of dilaton
 and axion hair for Kerr-Newman black holes [\asena -\cakaolb ]
combines all
of these scenarios\foot{For reviews of these and related developments,
see [\ghor -\asen ] . }.  There have also been recent studies of the
more physically interesting case of a massive dilaton coupled to an
Abelian charged black hole [\hohob -\grha ]. It is widely believed (but
 not required) that the dilaton acquires a mass when SUSY is broken: a
precisely massless dilaton violates the equivalence principle
[\shtrwi,\hohob ]
, and the dilaton cannot have a mass with SUSY intact. Since SUSY is
broken at low energies in any event, it seems essential that all of
the above scenarios be reexamined with a massive dilaton, though the
details of the SUSY-breaking mechanism and the dilaton potential are
not yet well understood.

The convergence of these separate efforts in black hole physics was
inevitable. A natural question to ask is whether non-Abelian gauge
fields in the low-energy string context lead to black holes with
primary hair, and if so, whether the sphaleron nature of previous
non-Abelian solutions is modified enough by the ``stringy'' scalar
fields to yield stable solutions. It was the desire to answer these
 questions, as well as to explore more general black hole solutions
to what could be {\it the} physically relevant theory, which motivated
the present work. While this paper was being completed, however, we
became aware of recent work in Einstein-Yang-Mills-Dilaton (EYMD)
theory [\lamaa -\bizob ,\dogaa -\dogab ]
which in part grew out of solutions to the Yang-Mills-Dilaton (YMD)
 system [\lamac -\bizoc ]. Though these efforts involve strictly
massless dilatons, some of our numerical results overlap with those of
 [\dogaa ] in which the authors examine a special case of the more
 general dilaton coupling $\gamma $ explored in [\lamaa -\lamab ]. We
draw comparisons to these numerical results wherever appropriate, and
discuss the implications of this recent body of work for the stability
of our solutions.

This paper is organized as follows. In Section 2, we introduce the
bosonic part of the low-energy heterotic string action, which we take
to first order in the inverse string tension after compactification to
$(3+1)$ dimensions. We specify the generic form of dilaton-axion
potential which arises when SUSY is broken by gaugino condensation, and
 obtain a simplified string-inspired theory by requiring spherical
symmetry and staticity, and by assuming that the characteristic
curvature of solutions is small compared to the Planck curvature. The
 spherically symmetric metric and $SU(2)$ connection ansatz are then
used to fully specify the theory, which is rewritten in terms of
dimensionless parameters and variables before the general field
equations are derived in Section 3. In Section 4, we classify all
possible non-Abelian solutions to the theory and ignore the embedded
Abelian solutions, which correspond to some of those discussed above
but with dilaton and axion masses included. Our analysis indicates that
 only two scenarios can admit solutions: a massive dilaton coupled to
 a single magnetic Yang-Mills degree of freedom,
and the full theory of a massive dilaton and massive axion
coupled to non-Abelian electric and magnetic fields. Though
 the latter theory is numerically intractable, we outline a possible
solution scenario before extensively analyzing and presenting
numerical regular and black hole solutions to the former theory in
Section 5. In the course of analyzing this theory, which we label EYMD
theory, we also note the equivalence of scaling arguments for the
existence of solutions [\hest ] and a judicious combination of the
field equations. In  Section 6, we speculate further on solutions to
the most general non-Abelian scenario, briefly address the issue of
stability, and offer our conclusions.

\nobreak

{\parindent=0pt\bf\chapter{ Preliminaries {}~\hfil~~\hfil~~\hfil~~
\hfil~~\hfil~~\hfil~~}}
{\parindent=0pt\bf Low-Energy String Action }

Our starting point is the bosonic part of the low-energy heterotic
string action [\grsl  ]
\def\K2{\kappa^2}\def\ap{\alpha^{\prime}}\def\Rgb{{\wdhat R}^2}
\def\mnl{\mu\nu\lambda}\def\mnrs{\mu\nu\rho\sigma}
\def\rlmn{\rho\lambda\mu\nu}
\def\brds{\left( D,s\right) }\def\brps{\left( \phi,s\right) }
$$\eqalign{
S=\gint\Biggl[ {R\over 2\K2} -{1\over 6} e^{-2\gamma D} H_{\mnl}
&H^{\mnl}-{1\over 2} \pd_{\mu} D\pd^{\mu} D -2V\brds \cr
-&{\ap\over 16 \K2}e^{-\gamma D}\left( 2g^2\fmni\fimn -\Rgb\right)
\Biggr]\cr}
\eqn\acta
$$
which is expressed in the Einstein frame for a metric with signature
$(-\ +\ +\ +)$. The action has been expanded to first order in the
inverse string tension $\alpha^{\prime} = 2\kappa^2 /g^2$, where $g$
is the gauge coupling for the Yang-Mills  (YM) curvature
$F= dA+gA\wedge A$
and $\kappa^2 = 8\pi G$. $H_{\mu\nu\lambda}$ is the field strength
tensor associated with the three-form
\def\olcs{\Omega_{3L}}\def\oymcs{\Omega_{3Y}}
\def\Trl{{\rm Tr}}\def\trg{{\rm tr}}
$$
H=dB+{\ap\over 8\kappa}\left(\olcs -g^2\oymcs\right)
\eqn\hform
$$
where $B$ is the two-form potential in the gravitational
supersymmetric multiplet.
$\Omega_{3L}$ and $\Omega_{3Y}$ are the Lorentz Chern-Simons (LCS)
 and  Yang-Mills Chern-Simons (YMCS) three-forms
$$\eqalignno{ \olcs &=\Trl \left[ \omega\wedge R-{1\over 3}
\omega\wedge\omega\wedge\omega\right] &\eqname\lcs\cr
\oymcs &=\trg\left[ A\wedge F-{1\over 3}
A\wedge A\wedge A\right] &\eqname\ymcs\cr}
$$
which arise in string theory in order to remove gauge and
gravitational anomalies.
Here $\trg $ and $\Trl $ denote trace over the suppressed
 gauge and Lorentz indices,
respectively, and the normalization for $\Omega_{3Y}$ is chosen for
gauge generators satisfying $\trg ( T^i T^j)=-2\delta^{ij}$. $R$ in
$\Omega_{3L}$ is not the first curvature scalar $R_{\mu\nu}g^{\mu\nu}$
that appears in the action; it is the curvature two-form
$$
R_{\mu\nu}=d\omega_{\mu\nu} +\omega_{\mu}{}^{\alpha} \wedge
\omega_{\alpha\nu}
\eqn\rcurv
$$
where $\omega_{a\mu\nu}= (e_{\mu})^{b}\nabla_a (e_{\nu})_{b}$ is the
spin connection for the tetrad $(e_{\mu})_{b}$. The other gravitational
 scalar appearing in \acta ~is the Gauss-Bonnet (GB)
curvature combination
$$
\Rgb = R_{\mnrs}R^{\mnrs}-4R_{\mu\nu}R^{\mu\nu}+R^2 ,
\eqn\gb
$$
which also helps to cancel anomalies and
is second-order in derivatives of $g_{\mu\nu}$. The dilaton field
$D$ couples to other fields through exponentials with coupling strength
$\gamma $ and has a self-interaction $V$ whose form will be specified
below.The normalization of $V$ has been chosen to accomodate a choice
 of coupling and a field rescaling: we take $\gamma =\sqrt{2}\kappa$
and define the dimensionless dilaton field
$\phi\equiv \kappa D /\sqrt{2}$, so
that the dilaton kinetic and potential terms assume the form
$$
{1\over 2\K2}\biggl( -2\pd_{\mu}\phi\pd^{\mu}\phi-4\K2 V\brps\biggr)
\eqn\dilkv
$$
It is important to note that $\gamma $ is the only coupling
parameter that we fix in
our analysis; all others (including the $\kappa^2$ factor now
appearing in front of $V$) will be absorbed in the definition of
other dimensionless fields and parameters.

The field $s$ appearing in $V$ is the dimensionless pseudoscalar
Kalb-Ramond axion, the only truly dynamical mode of the three-form
field which we introduce via
$$
H_{\mnl} ={1\over 2\kappa} e^{4\phi}\epsilon_{\mnl\sigma}
\pd^{\sigma}s .\eqn\sdef
$$
With this relation and the dual of the Bianchi identity
$$
dH={\ap\over 8\kappa}\left[ \Trl \left( R\wedge R\right)
-g^2 \trg\left( F\wedge F\right)\right]
\eqn\hbianc
$$
which follows from \hform , we can express the three-form field
strength as a sum of axion kinetic and topological current
contributions
$$
-{1\over 6} e^{-4\phi} H_{\mnl}H^{\mnl}= {1\over 2\K2}\left[
 -{1\over 2} e^{4\phi}\pd_{\mu} s\pd^{\mu} s
-2\kappa \nabla_{\mu} \left({}^*H\right)^{\mu}\right]
\eqn\hsub
$$
where $\kappa \nabla_{\mu} \left({}^*H\right)^{\mu}=
(\alpha^{\prime} /8)
\left[ \nabla_{\mu} \left({}^*\Omega_{3L}\right)^{\mu}
-g^2 \nabla_{\mu} \left({}^*\Omega_{3Y}\right)^{\mu}\right]$
is comprised of the four-divergence of two topological currents
$$\eqalignno{
\nabla_{\mu} \left( {}^*\Omega_{3L}\right)^{\mu} &=
-{1\over 2}\nabla_{\rho} \Trl  \Biggl[ \left( \omega_{[\lambda}
R_{\mu\nu ]} -{2\over 3} \omega_{[\mu}\omega_{[\nu}\omega_{\lambda ]]}
\right) \epsilon^{\rlmn} \Biggr] &\eqname\topcurl\cr
\nabla_{\mu} \left({}^*\Omega_{3Y}\right)^{\mu} &=
-{1\over 2}\nabla_{\rho} \trg \Biggl[ \left( A_{[\lambda}
F_{\mu\nu ]} -{2\over 3} g A_{[\mu}A_{[\nu}A_{\lambda ]]}
\right) \epsilon^{\rlmn} \Biggr] &\eqname\topcury\cr}
$$
which can also be expressed $-\Trl \left( R_{\mu\nu}\rtil^{\mu\nu}
\right) /2$ and $-tr\left( F_{\mu\nu}\ftil^{\mu\nu}\right) /2$,
respectively. With this replacement, the action becomes
$$\eqalign{
S=&{1\over 2\K2}\gint\Biggl[
R-{1\over 2} e^{4\phi}\pd_{\mu} s\pd^{\mu} s
-2\pd_{\mu}\phi\pd^{\mu}\phi-4\K2 V\brps \cr
&+{\ap\over 4}\left( {1\over 4} s\ \epsilon^{\rlmn}\left[
R_{\alpha\beta\rho\lambda}R_{\mu\nu}{}^{\alpha\beta}+2g^2
F^{(i)}_{\rho\lambda}\fmni\right] -{1\over 2}e^{-2\phi}\left[
2g^2\fmni\fimn -\Rgb\right]\right)\Biggr] \cr}
\eqn\actb
$$
The above equation is a general expression for the low-energy
heterotic string action to first order in the inverse string tension
for $\gamma =\sqrt{2}\kappa$. We now briefly examine some features of
this form of the action while arriving at some useful simplifications.

Note that when $V=0$, all sources for the dilaton and axion fields
are $\cal{O}(\alpha^{\prime} )$, so the fields themselves are first
order in $\alpha^{\prime}$. Furthermore, the sources are comprised of
gauge field and higher-derivative curvature combinations on an equal
footing. Even for fixed $s$, for which the topological current terms
arising from the YMCS and LCS three-forms contribute nothing to the
equations of motion, we must include the GB term ${\wdhat R}^2$ if
 we are to account for the gauge field strength. {}From this
perspective, the Reissner-Nordstrom solution (corresponding here to
the Abelian sector of some non-Abelian field strength and fixed
dilaton field) should for example be viewed as an
$\cal{O}(\alpha^{\prime} )$
correction to the Schwarzschild solution, subject to
 GB curvature corrections at the same order [\cakaolb ].
The gravitational effect of such curvature terms for fixed $\phi $
has been examined in
[\bodea -\wett ], although in $d=4$ the GB contribution enters purely
as a boundary term and can be ignored. The inclusion of the dilaton,
however, introduces an ${\wdhat R}^2$ source term in the
dilaton field equation even in $d=4$, and such scenarios have also
been studied [\bodeb -\kaol ]. Several authors have neglected the GB
curvature contribution in their investigations of the dilaton
while consistently keeping the gauge field strength source [\gahost
-\shtrwi ]
by considering solutions whose mass scale is large compared to
the Planck mass. In some circumstances, such as the
extremal limit of charged dilaton black holes [\gahost,\koyo -\gibb],
one can satisfy the mass scale assumption but introduce a different
inconsistency:  in this regime,  $\alpha^{\prime}$ is necessarily
large, so the dropping of higher order terms
in the effective string action is no longer justified [\prscshtrwi ].
Mindful of these concerns, we neglect the ${\wdhat R}^2$ term in the
action by assuming that the mass of solutions is large relative to
the Planck scale, but that $\alpha^{\prime}$ is small enough for
\actb ~to remain reliable.

For a dynamical axion $s$ with or without the dilaton, we again
encounter higher-order curvature and gauge field source terms. For
spacetimes with rotation, the LCS combination gives
 nontrivial contributions to the dilaton and axion equations of
motion, and analytical solutions for dilaton and axion hair outside
Kerr [\cdkoa -\mist ] and Kerr-Newman [\asena -\cakaolb ] black holes
have been obtained. For the 4-d Schwarzschild spacetime, spacetimes
related by a conformal transformation, or any 4-d spacetime with a
maximally symmetric 2-d
subspace, the LCS three-form either vanishes or is exact [\cdkob ].
Thus for the static, spherically symmetric spacetime we investigate
below, the remaining ${\cal O}\left( \alpha^{\prime} R^2\right)$
term in the action can be ignored.

The inclusion of the potential $V$ introduces an additional mass
scale into the problem, so that the dilaton and axion need not be
$\cal{O}(\alpha^{\prime} )$. None of the other preceeding observations
are qualitatively altered by its inclusion, but in choosing a
potential our discussion must move from generic features of heterotic
string theory (in static, spherically symmetric $(3+1)$ dimensional
 spacetime) to a more specific
model. Before doing so, we summarize the simplifications outlined
 above by rewriting the action
$$\eqalign{
S=&{1\over 2\K2}\gint\Biggl[
R-{1\over 2} e^{4\phi}\pd_{\mu} s\pd^{\mu} s
-2\pd_{\mu}\phi\pd^{\mu}\phi-4\K2 V\brps \cr
&+{\ap\over 4}\left( g^2\epsilon^{\rlmn}\pd_{\rho} s\left[
A^{(i)}_{\lambda}F^{(i)}_{\mu\nu} -{1\over 3} g\epsilon_{ijk}
A^{(i)}_{\mu}A^{(j)}_{\nu}A^{(k)}_{\lambda}\right] -e^{-2\phi}
g^2\fmni\fimn\right)\Biggr] \cr}
\eqn\actc
$$
where we have used the fact that $F_{\mu\nu }\ftil^{\mu\nu }$
is a 4-divergence ( eq.\topcury ~)
in a topologically trivial spacetime to recast the axion-gauge
 field coupling in a more convenient form.

{\parindent=0pt\bf Dilaton-Axion Potential}

We choose a potential of the form which arises
when supersymmetry is broken by gaugino condensation in the
hidden sector of the theory [\drsw ]
\def\alo{\alpha_{\infty}}\def\al{\alpha}
$$
V\brps = \mu^4 {\alo\over \al}\left[
1+ {\left(\al+1 \right)^2\over \left(\alo+1\right)^2}
e^{-\left(\al-\alo \right)}
-2{\left(\al+1 \right)\over \left(\alo+1\right)}
e^{-\left(\al-\alo \right) /2}
\cos \left({3\over 2b_0} s\right)\right]
\eqn\dilpot
$$
where $\alpha \equiv 3\exp(-2\phi)/b_{0}$, $\alpha_{\infty}$
corresponds to the
dilaton field at a potential minimum, and $b_{0}$ is determined by
the one-loop $\beta$-function of $Q$, the subgroup of the hidden
sector gauge group which precipitates supersymmetry breaking. We take
$Q$ to be the entire hidden sector gauge group that arises in these
scenarios, $E_8^{\prime}$, for which $b_{0}=90/(16\pi^2 )$. The
parameter
$\mu$ is a scale related to the vacuum expectation values of the
gaugino pair $\chi\overline\chi$ and the three-form $H_{m n p}$, where
$m$, $n$, and $p$ are indices on the internal compact manifold
$K$ only; we treat it here as a free parameter. With the axion field
set to its vacuum value, $s_{\infty}=b_{0}(4\pi n)/3$ for integer $n$,
this potential has been used by some authors to investigate inflation
 and cosmology in the context of superstring theories
(see e.g. [\kaol, \mapa -\kaolb ]). A plot of $V$ for
$s=s_{\infty}$ and $\phi_{\infty}=0$ is shown in fig.~1a; it has a
minimum at $\phi =\phi_{\infty}$ and achieves a local maximum at
$\phi <\phi_{\infty}$ before $V\rightarrow 0$ for
$\phi\rightarrow -\infty$.

{\parindent=0pt\bf Metric}

We parametrize the metric for a static, spherically symmetric
spacetime as
$$
ds^2=-T^{-2}\left( r\right) dt^2 + R^{2}\left( r\right) dr^2 +r^2\dom
,\eqn\meta
$$
where $\rr\equiv\left( 1-2G\mr/r\right)^{-1/2}$, $\mr$ is the total
mass-energy within the radius $r$, and we have set $c=1$.
To describe black hole solutions, we define $\delta\equiv\drr$ and
rewrite \meta ~as
$$
ds^2=-\R2\edr dt^2 +\R2^{-1} dr^2 +r^2\dom .\eqn\metbh
$$
Regularity at the origin requires
$ T\br0 < \infty$ and $R^{\prime}\br0 , T^{\prime}\br0 =0$ ,
 while regularity at the event horizon at $ r=r_{h}$ is satisfied by
$m\brh = r_{h}/2G$ and $\delta\brh <\infty$. We also impose
the condition of asymptotic flatness, which implies
$\rr,\tr\rightarrow 1$  as  $r\rightarrow \infty$ or, equivalently,
 $\rr\rightarrow 1$ and $\dr\rightarrow 0$.
By exploiting the freedom to rescale the time coordinate, however,
we can make the boundary conditions more suitable for integrating
the Einstein equations. Rather than requiring $T_{0}\equiv T\br0$ and
$\delta_{0}\equiv\delta\brh$ as initial conditions, consider
rescaling $t$ such that
$$
\eqalign{
R\br0 = 1, \ \ \ \ \ &T\br0 = 1\cr
 m\brh =r_{h}/2G,\ \ \ \ &\delta\brh =0  .\cr}\eqn\tdelc
$$
The condition of asymptotic flatness then translates into
$T\brinf =1/T_{0}$ and $\delta\brinf = -\delta_{0}$, and
we can determine $T_{0}$ or $\delta_{0}$ from the
behavior of a solution as $r\rightarrow \infty$.

{\parindent=0pt\bf Gauge Connection and YM Curvature}

In this paper, we investigate the simplest choice of non-Abelian
gauge group, $SU(2)$.
The most general spherically symmetric $SU(2)$ connection
 can be written in the form [\wit]
$$
A={1\over g}\left[ a\tar\ dt +b\tar\ dr +\left( d\tat -\c1
\tap\right) d\theta +\left( \c1\tat +d\tap\right) \sin\theta d\varphi
\right],\eqn\conna
$$
where $g$ is the gauge coupling and
$\left(\tar,\tat,\tap\right)$ is the
antihermitian $su(2)$ basis
projected along the polar coordinate directions:
$\tar={\bf\wdhat r}\cdot{\bf \tau}$, etc., and
the matrices satisfy $\left[ \taa,\tab\right] =\eabc\tac $
(we deviate from the gauge generator normalization used above for
this section only). {}From \conna ~we note that
 $c$ and $d$ are in general dimensionless functions
of $r$ and $t$, while
$a$ and $b$ have dimension $[L]^{-1}$.
The connection has a residual gauge freedom
under transformations of the form $U=\exp\left[ \brt \tar\right]$,
where $\brt$ is an arbitrary  real function, which we can use
to set ${b}\equiv 0$ while preserving the form of the connection.
Because the field equation for $b$ becomes a potentially useful
 constraint equation after gauge fixing, we leave $b$ nonzero for now
and demand that the component functions of $A$ depend on $r$ only.

Following [\wit], we express the $c$ and $d$
degrees of freedom in the connection
in  complex scalar form,
$$
c\brr -id\brr = f\brr \exp i\beta\brr ,\eqn\cdscalar
$$
which will make the non-Abelian character of the system more
transparent. The YM curvature $F=dA+gA\wedge A$ for the connection
 \conna ~is then
\def\dtth{\wdhat dt}\def\bmbp{\left( b-\beta^{\prime}\right)}
$$\eqalign{
F= {1\over g }
\biggl[ af {T\over r} \dtth +\bmbp { f\over rR } \drh\biggr]\wedge
&\biggl[ \phantom{-}\left( \cos\beta \tat -\sin\beta\tap\right)\dth
+\left( \sin\beta \tat +\cos\beta\tap\right)\dph\biggr] \cr
+{1\over g }\biggl[{ f^{\prime}\over rR } \drh\biggr]\wedge
&\biggl[-\left( \sin\beta \tat +\cos\beta\tap\right)\dth
+ \left( \cos\beta \tat -\sin\beta\tap\right)\dph
\biggr] \cr
+{1\over g } &\biggl[ -a^{\prime} {T\over R} \tar \dtth\wedge\drh
+{1\over r^2 }\left( f^2 -1\right) \tar \dth\wedge\dph\biggr]
 \cr}\eqn\fcurv
$$
which we have expressed in a convenient orthonormal tetrad basis,
$$
\dtth\equiv {1\over T} dt,\>\>\>\>\>\> \drh\equiv R dr ,\>\>\>\>\>\>
 \dth\equiv rd\theta ,
\>\>\>\>\>\> \dph\equiv r\sin\theta d\varphi .
\eqn\tetrad
$$
Note that the dependence of $F$ and $A$ on the gauge coupling
 indicates that $g$ only enters the simplified form of the action
\actc ~through the inverse string tension $\alpha^{\prime}$.

\nobreak

{\parindent=0pt\bf\chapter{ General Field Equations{}~\hfil~~
\hfil~~\hfil~~\hfil~~\hfil~~\hfil~~}}

With the choice of ansatz \conna~ for the gauge connection and
\dilpot ~for the dilaton-axion potential, we are in a position
to express the action in terms of the axion, dilaton, metric, and
gauge degrees of freedom. Before proceeding, we relax the condition
$c=1$ and examine the dimensionful quantities in the action in order
to cast our equations of motion in dimensionless form.

Noting that  $[g]=[T][M]^{-1/2}[L]^{-3/2}$,
 $[\mu^2]=[T]^{-1}[M]^{1/2}[L]^{-1/2}$, and that
$\kappa =\sqrt{ 8\pi G} /c^2 $ has dimensions
 $[\kappa]=[T][M]^{-1/2}[L]^{-1/2}$,  we observe that the parameters
appearing explicitly in the action have dimensions
$$
\left[\ap\right]=\left[ \K2 /g^2\right]=[L]^2 ,\>\>\>\>\>\>\>\>
\left[ \K2 \mu^4\right]=[L]^{-2} ,
\eqn\dimparam
$$
and that from \actb ~all terms in $2\kappa^2\lag$ have dimensions
$[L]^{-2}$. Thus, if we factor $(\alpha^{\prime}/4)^{-1}$
out of $2\kappa^2\lag$, define \def\gt{{\wdhat g}}
$\gt \equiv 2/\sqrt{\alpha^{\prime}}$,
 and define the dimensionless quantities
\def\ahat{{\wdhat a}}\def\bhat{{\wdhat b}}
$$
{\wdhat \mu}^2\equiv \K2\mu^2 /\gt ,\>\>\>\>\>\>
{\wdhat r}\equiv \gt r, \>\>\>\>\>\>
\ahat\equiv a/\gt,\>\>\>\>\>\> \bhat\equiv b/\gt,
\eqn\dimquant
$$
then $\detg (2\kappa^2\lag )$ can be written purely in terms of
dimensionless fields and parameters. To do this explicitly, we also
define a dimensionless mass-energy function $ \mhat $ based on
the metric \metbh :
$${1\over R^2\brr} = \left( 1-{2 G m\over c^2 r}\right) =
\left( 1-{2 \mhat\over \rhat}\right)
\>\>\>\> {\rm for} \>\>\>\>
{\mhat\left(\rhat\right)\equiv \gt G\ m\left( gr\right) /c^2 }.
\eqn\mhatdef
$$
Expressing the curvature scalar $R_{\mu\nu}g^{\mu\nu}$ in terms of
the metric functions $R(\mhat,\rhat )$ and $T(\rhat )$, we find the
following expression for the gravitational and
matter action of our static, spherically symmetric system:
$$\eqalignno{
S_{G}=& {c^4\over \gt G}\int dt d\rhat \Biggl[
{1\over 2}\left( {1\over R^2} -1\right) \rhat {d\over d\rhat}\left(
{R\over T}\right) \Biggr]
={c^4\over \gt G}\int dt d\rhat \Biggl[
-\mhat {d\over d\rhat}\left(
e^{-\delta} \right)   \Biggr]
&\eqname\actgr\cr
S_{M}=& {c^4\over \gt G}\int dt d\rhat \Biggl[
-\left( {1\over 8} e^{4\phi} {\left( \rhat s^{\prime}\right)^2
\over R^2}+{1\over 2}{\left( \rhat \phi^{\prime}\right)^2\over R^2}
+\rhat^2 {\wdhat V}\brps\right) {R\over T} -s^{\prime}\ahat
\left( f^2-1\right) &\eqname\acte\cr
&+e^{-2\phi}\left( T^2\left[ {1\over 2}  {\left( \rhat \ahat^{\prime}
\right)^2\over R^2} +f^2\ahat^2\right] -{1\over R^2}\left[
f^{\prime 2} +  f^2\left( \bhat-\beta^{\prime}\right)^2\right]
-{1\over 2 \rhat^2}\left( 1-f^2\right)^2\right) {R\over T}\Biggr]
&\cr}
$$
where the prime denotes derivative with respect to $\rhat$ and
${\wdhat V}\equiv \K2 V$.
The solutions to the dimensionless field equations obtained from this
action give us solutions for any ${\wdhat g}>0$ ( or
$0< \alpha^{\prime}<\infty$)  through scaling relations,
\def\bgrr{\left( \gt r\right)}
$$
a_{\gt}\brr =\gt\ahat\bgrr ,  \>\>\>\>\>\> m_{\gt}\brr =
{c^2\over \gt G} \mhat\bgrr , \>\>\>\>\>\>
{\cal F}_{\gt}\brr = {\cal F}\bgrr ,
\eqn\scalesoln
$$
where ${\cal F}$ denotes any of the functions  $\{ s,\ \phi,\ f,
\ \beta,\ R,\ T,\ \delta\}$ and we have ignored $b$ since it will be
eliminated by gauge fixing. Hence the radial
structure of solutions for a given value of ${\wdhat g}$ is the same
as that obtained from \acte , but it occurs at a physical radius
$r=\rhat/ {\wdhat  g}$ with physical scales given by \scalesoln ~and
$\mu^2 = {\wdhat g} {\wdhat \mu}^2/\kappa$. For notational simplicity
throughout the remainder of the paper, we drop the carets on
dimensionless quantities with the understanding that everything is
now dimensionless unless otherwise specified.

By varying \acte ~with respect to the  fields, we obtain the
dimensionless, static field equations
$$\eqalignno{
&{d\over dr}\left({r^2\phi^{\prime}\over RT}\right)
-{\pd V\brps \over \pd\phi} r^2 {R\over T}
-{1\over 2} e^{4\phi} {\left( r s^{\prime}\right)^2\over R T}
&\eqname\phieqn\cr
& +2\ep\left( {1\over R^2}\left[ f^{\prime 2}+f^2
\left( \beta^{\prime}-b\right)^2 \right]+
{\left( 1-f^2\right)^2\over 2 r^2}
-T^2\left[ {1\over 2}  {\left( r a^{\prime}
\right)^2\over R^2} +f^2 a^2\right] \right) {R\over T} =0 &\cr
&{d\over dr}\left( {1\over 4}{ e^{4\phi}\over  RT}r^2 s^{\prime}\right)
-{\pd V\brps \over \pd s} r^2 {R\over T}
- {d\over dr}\left( a\left[ 1-f^2\right]\right) =0
&\eqname\seqn\cr
&{d\over dr}\left( \ep {T\over R} r^2a^{\prime}\right)
-2\ep f^2 a RT-s^{\prime}\left( 1-f^2\right) =0 &\eqname\aeqn\cr
&{d\over dr}\left( {\ep \over RT}f^{\prime}\right) +
\ep\left({\left[ 1-f^2\right]\over r^2} f
-{\left[ \beta^{\prime} -b\right]^2
\over R^2} f + T^2 a^2 f\right){R\over T} -s^{\prime}af
=0 &\eqname\feqn\cr
&{d\over dr}\left( {\ep\over RT} f^2\left[\beta^{\prime}-b\right]
\right) =0, &\eqname\beteqn\cr
}
$$
and the constraint equation
$$
{\ep\over RT}f^2\left(\beta^{\prime}-b\right)=0 .\eqn\bconstr
$$
Even without using the remaining gauge freedom to set
$b\equiv 0$, we find that $f^2\left(\beta^{\prime}-b\right) $
disappears from the field equations. The constraint equation with
gauge fixing does give us additional information, however: it implies
that our gauge choice eliminates an additional degree
of freedom ($c$ and $d$ are related by a multiplicative constant)
when we insist $f\not\equiv 0$, a criterion for non-Abelian solutions.
In the Abelian sector $f\equiv 0$, it and the
$\beta $ field equation indicate that
$\beta^{\prime} $ is an arbitrary function of radius, which
reflects the fact that $c\equiv d\equiv 0$ in \cdscalar ~entirely
eliminates the the need for the complex scalar phase. In either case,
we can now eliminate $f^2\left(\beta^{\prime}-b\right) $ from our
analysis.

To obtain the Einstein equations, we can either utilize the
energy-momentum tensor,
$$\eqalignno{
{\K2\over \gt^2} T_{\mu\nu} = &\ep\left[ 2F_{\mu\gamma}^{(i)}
F_{\nu}^{(i)}{}^{\gamma} -{1\over 2}
g_{\mu\nu}\fmni\fimn \right] +{1\over 2} e^{4\phi}\pd_{\mu} s
\pd_{\nu}s &\cr
 &+2\pd_{\mu}\phi\pd_{\nu}\phi -g_{\mu\nu}\left[
{1\over 4} e^{4\phi}\pd_{\rho} s\pd^{\rho}s
+\pd_{\rho}\phi\pd^{\rho}\phi +2 V\brps \right] ,
&\eqname\tabeymh\cr}
$$
or use the explicit dependence of \actgr ~and \acte ~on
some pair of independent metric functions ($m$ and $\delta$ will do
as well as $R$ and $T$) to derive the gravitational equations
 directly. By whatever route, we find the
 $(tt)$ and $(rr)$ Einstein equations can be expressed in the form
\def\Rr2{\left( 1-{2 m\over r}\right)}
$$\eqalignno{
m^{\prime} =\ep &\left[ \Rr2 f^{\prime 2}
+{\ww1^2\over 2r^2} +
T^2\left( {1\over 2} \Rr2 {\left( r a^{\prime}
\right)^2} +f^2 a^2\right)\right] &\cr
&+\Rr2\left[ {1\over 8} e^{4\phi} {\left( r s^{\prime}\right)^2}
+{1\over 2}{\left( r \phi^{\prime}\right)^2}\right]
 +V\brps r^2 &\eqname\meqn\cr
r\Rr2{T^{\prime}\over T} =
\ep &\left[ -\Rr2 f^{\prime 2}
+{\ww1^2\over 2r^2} +
T^2\left( {1\over 2} \Rr2 {\left( r a^{\prime}
\right)^2} -f^2 a^2\right)\right] &\cr
&\phantom{ f }\; -\Rr2\left[
{1\over 8} e^{4\phi} {\left( r s^{\prime}\right)^2}
+{1\over 2}{\left( r \phi^{\prime}\right)^2}\right]
 +V\brps r^2 -{m\over r} .&\eqname\teqn\cr }
$$
For black hole solutions, we replace the $T^{\prime}$ equation with
$$
\delta^{\prime} =-{2\over r}\left[ \ep \left[ f^{\prime 2} +
\Rr2^{-2} e^{2\delta} f^2 a^2\right] +
{1\over 8} e^{4\phi} {\left( r s^{\prime}\right)^2}
+{1\over 2}{\left( r \phi^{\prime}\right)^2}\right]  ,\eqn\deqn
$$
which indicates that $\delta $  monotonically decreases
with radius. {}From this fact and the
boundary condition $\delta\brinf =0$, we can conclude that
$\delta\ge 0$ and thus $T\brr\ge R\brr \ge 1$.
Another metric relation
useful for predicting solution properties is
$$\eqalign{
{d\over dr}\left[{r^2\over R}\left( {1\over T}\right)^{\prime}\right]
= &\left[ 2 \ep\left[ \Rr2 f^{\prime 2} +{\ww1^2\over 2 r^2}\right]-
2V\brps r^2\right]{R\over T}\cr
 &+2 \ep\left[{1\over 2} \Rr2 {\left( r a^{\prime}
\right)^2} +f^2 a^2 \right] RT\cr},\eqn\tcon
$$
which may be obtained by combining the field equations. In preparing
the  equations for integration, we will also require the
relation
$$
r^2 {T\over R}
{d\over dr}\left({1\over RT}\right) =\left( 2m-\ep
\left[ r\Rr2 \left( ra^{\prime}\right)^2 T^2 + {\ww1^2\over r}
\right] -2V\brps r^3\right)
\eqn\deriv
$$
Finally, in classifying the possible solutions of this system of
equations, it is useful to rewrite \aeqn ~as
$$
{1\over 2}{d\over dr}\left[\ep {T\over R} r^2 \left(a^2
\right)^{\prime}\right]=
\ep {T\over R} \left( ra^{\prime}\right)^2+
2\ep  f^2 a^2  RT +s^{\prime}\ww1 a.\eqn\aeqnb
$$
The positivity properties of the right-hand side of this equation,
coupled with boundary conditions, can be applied to establish
no-dyon results in the non-Abelian sector analogous to those for
EYM [\erga ] and EYMH [\grmaon ] theories.

\nobreak

{\parindent=0pt\bf\chapter{Taxonomy of Non-Abelian Solutions
{}~\hfil~~\hfil~~\hfil~~\hfil~~\hfil~~}}

In previous studies of $SU(2)$ gauge theories coupled to Einstein
gravity, many authors have noted that $f\equiv 0$ in the gauge
connection ansatz gives a theory with Coulombic (i.e. $U(1)$)
magnetic and electric charges. As discussed in the introduction,
a great deal of work has recently been done on the
corresponding Abelian sector of the low-energy heterotic string
action. Though the addition of the SUSY-breaking potential \dilpot
{}~could provide
interesting new features for some Abelian sector solutions (see,
for example, [\hohob -\grha ]), we choose not  to explore them here.
Our focus is the non-Abelian sector of the theory \actc, and in this
section we consider the various possibilities for static, spherically
symmetric solutions with $f\not\equiv 0$.

The most general class of possible non-Abelian solutions to the field
equations corresponds to $\{ f,\ a,\ s,\ \phi \}$ behaving as
nontrivial functions of radius. By studying the asymptotic behavior
of the field equations, we find that the condition $f\not\equiv 0$
requires
$f^2\brinf =1$, which further implies $a\brinf =0$ via \aeqn -\feqn .
The decay of $f$ and $a$ toward these asymptotic values is also too
rapid to give a net electric or magnetic charge, so such solutions are
not dyons in the usual sense but do possess both electric and
magnetic fields.
The asymptotic values of $\phi$ and $s$ are fixed by $V$; as we
observed in the discussion of the dilaton-axion potential, we are
free to choose $\phi_{\infty}$ but the axion must assume the value
$s_{\infty}=b_{0}(4\pi n)/3$ for some integer $n$. At the origin or
horizon $r_h$ of a black hole solution, we must choose $\{
f^{\prime\prime}\br0, \ a^{\prime}\br0,\ s\br0, \ \phi\br0 \}$ and
$\{ f\brh,\ a^{\prime}\brh,\ s\brh,\ \phi\brh\}$, respectively, to
integrate the equations of motion; in either case finite energy
density restrictions force the initial value of $a$ to vanish.
Exploring a system with four integration parameters is impractical
numerically, but we can demonstrate analytically that solutions are
not forbidden to exist, and we can anticipate some of the properties
of potential solutions.
The analysis of the general case is best done, however, after the
simpler possibilities are surveyed.
To describe the less general non-Abelian solutions, we reduce the
number of integration parameters by ansatzing
each of the fields $\{ f,\ a,\ s,\ \phi \}$ to an appropriate
constant value in turn. We will find that only the $a\equiv constant$
and $s\equiv constant $ cases admit nontrivial solutions, though the
$\phi\equiv constant$ ansatz provides much insight into the
characteristics of the most general non-Abelian solutions.

{\parindent=0pt\bf ${\bf f\equiv constant}$}

The asymptotic behavior of the field equations restrict the acceptable
values to $f^2\equiv 1$, and the alternate version of the $a$ equation
\aeqnb ~can then be used to establish $a\equiv 0$. The proof relies
on the fact that the right-hand side of \aeqnb ~is nonnegative for
$f^2\equiv 1$, so that $a^2$ is strictly increasing beyond the initial
radius for nontrivial
solutions. The boundary conditions $a\br0=a\brh=a\brinf=0$,
however, require that $a^2$ decrease toward $a^2=0$ asymptotically.
It then follows that the only solution consistent with the $a$ field
equation and the boundary conditions is the trivial solution
$a\equiv 0$, and we are left with a theory of two gravitating scalar
fields. Through the metric relation \tcon ~and the same line of
reasoning used to establish $a\equiv 0$, we can now show that only
trivial solutions follow
from $f\equiv constant$. Since the right-hand side of that relation
is never positive, nontrivial solutions obey the condition
$T^{\prime}>0$ beyond the initial radius. Because we also noted in
the discussion of \deqn ~that $T\brr\ge T\brinf =1$, the only way to
reconcile monotonically increasing $T$ with the boundary conditions
is to require
 $V\left(\phi, s\right)\equiv 0$: only the trivial solution
$\phi\equiv\phi_{\infty}$, $s\equiv s_{\infty}$ is compatible with
both the boundary conditions and the field equations.

{\parindent=0pt\bf ${\bf a\equiv constant}$ or ${\bf s\equiv
constant}$: EYMD Theory}

By setting either $a$ or $s$ to a constant value, we arrive at a
theory with one gauge degree of freedom coupled to a massive dilaton
and Einstein
gravity, which we denote Einstein-Yang-Mills-Dilaton (EYMD) theory.
In the $s\equiv constant$ case, the potential requires $s\equiv
s_{\infty}$, and the $a$ equation \aeqnb ~with $s^{\prime}=0$
subsequently yields $a\equiv 0$ when we employ the reasoning
introduced in the
$f\equiv constant$ discussion. Alternately, if we take $a\equiv
constant$ the boundary conditions require $a\equiv 0$, and the field
equations then imply $s\equiv s_{\infty}$ for $f^2\not\equiv 1$. Note
that the right-hand side of the metric relation \tcon ~again involves
a $-V $ contribution, but now there are positive $f$-dependent terms
which make possible the decrease of $T\brr$ toward $T\brinf =1$
required for nontrivial solutions.
To obtain regular or black hole solutions to EYMD theory, we must fix
$\phi_{\infty}$ and choose the integration  parameters
$\{ f^{\prime\prime}\br0,\ \phi\br0 \}$  or $\{ f\brh, \ \phi\brh \}$,
respectively, such that the fields match the appropriate  boundary
conditions upon integration. We provide a detailed discussion of this
procedure, which is often referred to as a two-parameter ``shooting''
procedure, and obtain numerical solutions in the next section.

It is interesting to note that in the absence of the dilaton-axion
potential, solutions to EYMD theory are the most general non-Abelian
 solutions to the full field equations: dyonic non-Abelian solutions
and nontrivial axion solutions are prohibited when $V\equiv 0$.
The $s$ field equation \seqn ~in this case implies
$$
s^{\prime} = 4 \>{RT\over r^2} e^{-4\phi}\left[
a\ww1 + c_{s^{\prime}} \right],\eqn\seqnb
$$
where $c_{s^{\prime}}$ is a constant which acts asymptotically like a
Coulombic charge, giving a contribution to the total mass-energy
proportional to $c_{s^{\prime}}^2/r^2$ at large radius. Though
nonzero $c_{s^{\prime}}$ appears acceptable at large $r$, it violates
regularity of the metric at the origin for regular solutions and at
the event horizon for black hole solutions, since $m^{\prime}\br0$ and
 $\delta^{\prime}\brh$ diverge as $\left( r s^{\prime}\right)^2$
diverges. When \seqnb ~with $c_{s^{\prime}}=0$ is substituted into
the alternate form of the $a$ equation \aeqnb, the right-hand side is
nonnegative and $a\equiv 0$ by the arguments outlined above. It then
follows from \seqnb ~that $s\equiv constant$, the value of which is
no longer fixed by the minimum of $V$, and the general theory reduces
to EYMD theory for
a massless dilaton.  A closer look at the field equations, however,
reveals that  $a\equiv 0$ and $s\equiv constant$ also
follows from the weaker condition $(\pd V/\pd s)\equiv 0$: the no-dyon
result is a consequence of having an axion-independent potential,
rather than no potential at all. The possibility of non-Abelian
solutions other than the EYMD class relies crucially on the presence
of a massive (or at least self-interacting) axion field in the theory.

{\parindent=0pt\bf ${\bf \phi\equiv constant}$ }

At first glance, setting $\phi$ equal to its asymptotic value
$\phi_{\infty}$ appears to reduce the number of independent matter
fields from four to three. The exponential coupling of the dilaton to
the gauge field and axion kinetic terms in \phieqn ~instead make the
dilaton
field equation a nontrivial constraint which must be satisfied by
the remaining fields $\{ f,\ a,\ s\}$. Using this constraint and some
field redefinitions, it is possible to cast the resulting theory
in a form  which requires only two independent integration parameters.
Though the constraint equation appears to make the theory
 numerically tractable,
it is also responsible for the nonexistence of solutions: by
differentiating with respect to $r$, we find that the constraint is
incompatible with the $\phi\equiv constant$ field equations. Because
this theory shares important features with the most general
non-Abelian case, we
develop the two-parameter formulation and show explicitly that
solutions cannot exist. \def\ac{{\cal A}}\def\ec{{\cal E}}

The prospect of black hole solutions with nontrivial gauge degree of
freedom $a$ motivates us to define $\ac\equiv aRT$: based on
the expression \deqn ~for $\delta^{\prime}$, $a$ must vanish at least
as fast as $1/RT$ near the horizon for the $a^2$ term not to diverge
and violate regularity requirements.  It is also convenient to
introduce $\ec\equiv e^{\delta} a^{\prime}$ in lieu of $\ac^{\prime}$,
to which $\ec$ is related by
$$
\ec=\left( {\ac\over R^2}\right)^{\prime} -\delta^{\prime}
{\ac\over R^2}={1\over R^2}\left( \ac^{\prime}-\delta^{\prime}
\ac\right) -2\left({m\over r}\right)^{\prime}\ac .
\eqn\eceqn
$$
In terms of these new variables, the self-interaction and kinetic
contributions of $a$ to the action interchange roles; the $1/R^2$
factor which appears with $(a^{\prime})^2$ is absorbed by the
definition of $\ec$, and reappears in the $f^2a^2$ term:
$$
T^2\left( {1\over 2} \Rr2 {\left( r a^{\prime}
\right)^2} +f^2 a^2\right) ={1\over 2} r^2\ec^2 +{ f^2\ac^2\over R^2} .
\eqn\ecacsub
$$
This combination also appears in the Einstein equations \meqn -\teqn
{}~and the metric relation \tcon, which are unchangedapart from the
substitution of
\ecacsub ~and $\phi\equiv\phi_{\infty}$. The interchange of kinetic
and potential roles is sensible when we consider the expression for
$\delta^{\prime}$,
$$
\delta^{\prime} =-{2\over r}\left[ e^{-2\phi_{\infty}} \left[
f^{\prime 2} +f^2 \ac^2\right] +
{1\over 8} e^{4\phi_{\infty}} {\left( r s^{\prime}\right)^2}\right]  ,
\eqn\deqnphi
$$
which now involves only kinetic terms and is no longer explicitly
dependent on the metric functions.  Because the relation between
$\ec$ and $\ac^{\prime}$ depends explicitly on $\delta^{\prime}$, it
is convenient to use \deqnphi ~in place of the Einstein equation
\teqn ~when considering regular as well as black hole solutions.

Through the metric derivative equation \deriv, which becomes
$$
r^2 e^{\delta}
{d\over dr}\left({e^{-\delta}\over R^2}\right) = 2r\left( {m\over r}
-e^{-2\phi_{\infty}}
\left[ {1\over 2} r^2 \ec^2 + {\ww1^2\over 2 r^2}
\right] -V\left( \phi_{\infty},s\right) r^2\right)
\eqn\derivphi
$$
for this case, we can also express the axion and gauge field equations
in a form independent of either $T$ or $\delta$,
$$\eqalignno{
& {r^2\over R^2} s^{\prime\prime} +\left[ {2 r\over R^2}+
r^2 e^{\delta}
{d\over dr}\left( { e^{-\delta}\over  R^2}\right)\right] s^{\prime}
- 4 e^{-4\phi_{\infty}}r^2\left[{\pd V\brps \over \pd s}
\right]_{\phi=\phi_{\infty}}
 &\eqname\seqnp\cr
&\phantom{
{r^2\over R^2} s^{\prime\prime} +\left[ {2 r\over R^2}+
r^2 e^{\delta}
{d\over dr}\left( { e^{-\delta}\over  R^2}\right)\right] s^{\prime}
}
- 4 e^{-4\phi_{\infty}}\left[ \ec\ww1 -2ff^{\prime}
{\ac\over R^2}\right] =0 &\cr
& r^2 \ec^{\prime} + 2r\ec -2 f^2 \ac -e^{2\phi_{\infty}}
s^{\prime}\ww1 =0 &\eqname\aeqnp\cr
& {r^2\over R^2} f^{\prime\prime} +\left[
r^2 e^{\delta}
{d\over dr}\left( { e^{-\delta}\over  R^2}\right)\right] f^{\prime}
+\left[ \ww1 +r^2 {\ac\over R^2}\left( \ac -
e^{2\phi_{\infty}} s^{\prime}\right)\right] f=0 .&\eqname\feqnp\cr
 }
$$
The potential for $\phi\equiv \phi_{\infty}$ may be written in the
form
$$
V\left( \phi_{\infty},s\right) =
2\mu^4\left( 1- \cos\left[ {1\over 2}\left( {3\over b_0} s\right)
\right]\right) =
 4\mu^4 \sin^2\left[ {1\over 4}\left( {3\over b_0} s\right)\right] ,
\eqn\vpinf
$$
which is proportional to the derivative with respect to $\phi$ at
$\phi_{\infty}$:
$$
\left[{\pd V\brps \over \pd \phi} \right]_{\phi=\phi_{\infty}}
=A_{\phi_{\infty}} V\left( \phi_{\infty},s\right)\equiv
\left[ {\alo\left( \alo+1\right) +2\over
\left( \alo +1\right)}\right]V\left( \phi_{\infty},s\right) ,
\eqn\dvpinf
$$
where $\alo \equiv 3\exp(-2\phi_{\infty})/b_{0}$ and we have
introduced $A_{\phi_{\infty}}$ to denote the constant of
proportionality. This property of the potential allows us to
write the dilaton constraint equation in the virial-like form
$$e^{-2\phi_{\infty}}
\left({f^{\prime 2}\over R^2}
+{\ww1^2\over 2r^2}
-{1\over 2} r^2\ec^2 -{ f^2\ac^2\over R^2}\right)
-\left( {1\over 4} e^{4\phi_{\infty}} {\left( r s^{\prime}\right)^2
\over R^2} +{1\over 2} A_{\phi_{\infty}}
V\left( \phi_{\infty},s\right) r^2\right) =0.
\eqn\peqnp
$$
We can use this constraint to help integrate the system comprised of
the $m^{\prime}$ and $\delta^{\prime}$ equations, eqns.\seqnp -\feqnp,
and the $\ec$-$\ac^{\prime}$ relation \eceqn : it reduces the number
of integration parameters to two and provides a check as we integrate
the system.

There are reasons to expect nontrivial solutions to this
theory. We can formally integrate \seqnp ~to obtain an expression
analogous to \seqnb,
$$
{1\over 4} e^{4\phi_{\infty}} {\left( r s^{\prime}\right)^2
\over R^2} = s^{\prime}\ww1{\ac\over R^2} +e^{\delta} s^{\prime}
\left[ \int\nolimits_{\infty}^{r} d\rt \> \rt^2 {\pd V\left(
\phi_{\infty},s\right)\over \pd s} e^{-\delta\left( \rt\right)}
\right] ,\eqn\seqnpb
$$
which upon substitution into the alternate form of the $a$ equation
\aeqnb ~gives us
$$\eqalign{
{d\over dr}\left[ {e^{-\delta} \over R^2} e^{-2\phi_{\infty}} r^2
\ac\ec \right] &= 2 e^{-\delta}\left[e^{-2\phi_{\infty}}
\left({1\over 2} r^2\ec^2 +{ f^2\ac^2\over R^2}\right)
+{1\over 2} s^{\prime}\ww1{\ac\over R^2}\right] \cr
 &= 2 e^{-\delta}\left[e^{-2\phi_{\infty}}
\left({1\over 2} r^2\ec^2 +{ f^2\ac^2\over R^2}\right)
+{1\over 8} e^{4\phi_{\infty}} {\left( r s^{\prime}\right)^2
\over R^2}\right] \cr
&\phantom{ = 2 e^{-\delta}\left[e^{-2\phi_{\infty}}
\right]} -s^{\prime}\left[ \int\nolimits_{\infty}^{r} d\rt \> \rt^2
{\pd V\left( \phi_{\infty},s\right)\over\pd s} e^{-\delta\left( \
rt\right)}\right]  .\cr}
\eqn\aeqnpb
$$
In the two previous  cases considered, we used the manifest
positivity of the right-hand side of this equation to help establish
$\ac =\ec\equiv 0$ and $s\equiv constant$ as the only acceptable
solution. Now
we must contend with the final term in \aeqnpb, the sign and magnitude
of which depend on the details of the potential. For the pure axion
potential depicted in fig.~1b, we can imagine a solution scenario
in which $s$ increases from some initial value
$4\pi n ( b_0 /3) < s_{0} < 4\pi (n +1/2)( b_0 /3)$,
moves over the potential maximum at
$s_{{max}}= 4\pi (n +1/2)( b_0 /3)$, and comes to rest at the closest
degenerate minimum $s_{\infty}= 4\pi (n +1)( b_0 /3)$.
With $s^{\prime} \ge 0$ along the entire solution trajectory, the
sign of the final term in \aeqnpb ~is completely determined by the
integral factor. For $s_{{max}} < s<s_{\infty}$, the integrand is
negative-definite. For $s_{0} <s<s_{{max}}$, the sign of
$\pd V/\pd s$ changes
but the integrated contribution from $r>r(s_{{max}})$ initially
dominates. Since the monotonically increasing factor $r^2e^{-\delta}$
in the
integrand weights the $r>r(s_{{max}})$ contribution more heavily, and
$\left| \pd V/\pd s\right|$ is symmetric about $s=s_{{max}}$, we can
in fact conclude
$$
-s^{\prime}\left[ \int\nolimits_{\infty}^{r} d\rt \> \rt^2 {\pd V
\left( \phi_{\infty},s\right)\over\pd s} e^{-\delta\left( \rt\right)}
\right] \le 0 \eqn\sintsign
$$
for the entire trajectory, and the right-hand side of \aeqnpb ~is not
positive definite. In order that $a$ be nontrivial, however, the
magnitude of the contribution \sintsign ~must be large enough that
the integral of the right-hand side of \aeqnpb ~be negative as
$r\rightarrow\infty$, which from the original equation \aeqnb ~is
required for the asymptotic decrease of $a^2$ toward zero.
{}From \seqnpb, we must therefore have
$${1\over 4} e^{4\phi_{\infty}} {\left( r s^{\prime}\right)^2
\over R^2} -e^{\delta} s^{\prime}
\left[ \int\nolimits_{\infty}^{r} d\rt \> \rt^2 {\pd V\left(
\phi_{\infty},s\right)\over \pd s} e^{-\delta\left( \rt\right)}
\right]= s^{\prime}\ww1{\ac\over R^2} < 0
\eqn\seqnpc
$$
at least as $r\rightarrow \infty$,
from which we conclude $\ac\ww1 <0$ asymptotically and $\ac<0$
if $f^2$ approaches $f^2\brinf =1$ from below. We can infer some
general features of $f^2$ by rewriting the $f$ field equation as
$$
{1\over 2}e^{\delta}{d\over dr}\left[ { e^{-\delta}\over  R^2}
e^{-2\phi_{\infty}}\left( f^2\right)^{\prime}\right] =
e^{-2\phi_{\infty}} {f^{\prime 2}\over R^2} -
e^{-2\phi_{\infty}} \left[ \ww1 +r^2 {\ac\over R^2}\left( \ac -
e^{2\phi_{\infty}} s^{\prime}\right)\right] {f^2\over r^2}.
\eqn\feqnpb
$$
{}From this expression, we find that solutions must obey the inequality
$$
f^2\le 1+r^2 {\ac\over R^2}\left( \ac -
e^{2\phi_{\infty}} s^{\prime}\right) ,
\eqn\flimp
$$
when $\ac <0$  or  $\ac> e^{2\phi_{\infty}} s^{\prime}$,
since outside this region the right-hand side of \feqnpb ~is
positive-definite and $f^2$ cannot approach unity asymptotically.
For $0<\ac <e^{2\phi_{\infty}} s^{\prime}$, on the other hand, $f^2$
can exceed the bound in \flimp ~but must then approach $f^2\brinf $
monotonically, so the inequality simplifies to $f^2\le 1$. Further
implications follow from the
original form of the $f$ equation \feqnp, which indicates that
$f f^{\prime\prime} < 0$ when $f^{\prime} =0$ and \flimp ~is satisfied
regardless of the value of $\ac$: solutions can exhibit oscillations
about $f=0$ at finite radius,
but they cannot possess turning points once the inequality bound is
exceeded. Combining these observations, we can construct a simple
picture for the $s^{\prime}\ge 0$ trajectory that is consistent with
the field equations: $\ac \le 0$, with a single
extremum and increase toward $\ac\brinf =0$ as the negative
contribution \sintsign ~begins to dominate in \aeqnpb, and
$f^2\lsim 1$, with oscillations and nodes in $f$ possible. Had we
begun with
$4\pi (n +1/2)( b_0 /3) < s_{0} < 4\pi (n+1)( b_0 /3)$ and demanded
that $s$ decrease monotonically toward $s_{\infty}=4\pi n ( b_0 /3)$,
we would have found an identical picture with $\ac\rightarrow -\ac$.
Thus, in the presence of a sufficiently massive axion field,
non-Abelian dyonic solutions for the $\phi\equiv constant$ theory
appear possible.

Unfortunately, the constraint \peqnp ~that follows from the
$\phi\equiv constant$ ansatz is not consistent with the
rest of the field equations. It is not obvious from inspection
whether solutions to \seqnp -\feqnp, \eceqn, and the Einstein
equations satisfy \peqnp ~for some choice of the parameters
$\{ \mu, \phi_{\infty} \}$. Because this set of equations without the
constraint is sufficient to determine solutions, we should be able to
 verify \peqnp ~by utilizing the entire
set. When we take the derivative of \peqnp ~with respect to $r$ and
use the other field equations to eliminate higher derivatives, we
obtain the relation
$$\eqalign{
2 e^{-2\phi_{\infty}}\left(
{\left[ \ww1^2\right]^{\prime}\over 2r^2}  -
\left[ \left( f^2\right)^{\prime}{\ac^2\over R^2} +2\ac\ec f^2\right]
\right) -3 s^{\prime}\left( \ww1^{\prime}{\ac\over R^2}
+\ww1\ec\right) & { }\cr
+{2\over r}\left( e^{-2\phi_{\infty}}
{f^{\prime 2}}- e^{-2\phi_{\infty}} f^2\ac^2
-{1\over 4} e^{4\phi_{\infty}} \left( r s^{\prime}\right)^2 \right)
\left( {1\over R^2} +{r\over R^2}\left( \ln T\right)^{\prime}\right)
& { }\cr
-2\left(1+{1\over 4}A_{\phi_{\infty}}\right) r^2 {\pd
V\left( \phi_{\infty},s\right)\over \pd s}s^{\prime} + {2\over r}
\left( {1\over 4} e^{4\phi_{\infty}} {\left( r s^{\prime}\right)^2
\over R^2}-A_{\phi_{\infty}}
V\left( \phi_{\infty},s\right) r^2\right) &=0\cr
}\eqn\derivcon
$$
which should be satisfied along with \peqnp ~for all $r$. It appears
that no choice of $\{ \mu, \phi_{\infty} \}$ or further simplification
with the field equations can make \derivcon ~an identity for
nontrivial gauge and axion fields, and we conclude that the
$\phi\equiv constant$ case admits only trivial solutions.

{\parindent=0pt\bf General Non-Abelian Solutions}

The failure of the $\phi\equiv constant$ system to admit nontrivial
solutions is closely tied to the constraint created by a nondynamic
dilaton field. How does the situation change when we relax the
restriction that $\phi\equiv \phi_{\infty}$?

Exciting the dilaton degree of freedom restores the original form of
the potential \dilpot, but the expansion of the potential in powers of
\def\phat{{\wdhat \phi}}\def\ord{{\cal O}}
 $\phat\equiv \phi-\p0$ can be written in the form
$$\eqalign{
V\left( \phat ,s\right) =V\left( \phi_{\infty},s\right)
&\left[ 1+ {\left( \alo^2 +\alo +2\right)\over
\left( \alo +1\right) }\phat +
{\left( \alo^4 +\alo^2 +6\alo +4\right)\over 2
\left( \alo +1\right)^2 }\phat^2 + \ord\left( \phat^3\right)
\right] \cr
&\phantom{\left[ 1+ {\left( \alo^2 +\alo +2\right)\over
\left( \alo +1\right) }\phat\right]}
+\mu^4\alo^2\left( {\alo -1\over \alo +1}
\right)^2\phat^2 +\ord\left( \phat^3\right) .\cr }
\eqn\dilpotexp
$$
Thus $\pd V /\pd s$ deviates from the purely axionic form discussed
above, but only by a $\phat$-dependent scale factor. This scale
factor  breaks the symmetry of $\left| \pd V /\pd s\right|$
about $s=s_{{max}}$, but it is of order unity
when $\phi $ is near the minimum at $\phi_{\infty}$,
so the reasoning behind \sintsign ~for the $s^{\prime}\ge 0$ solution
scenario remains intact when the expansion \dilpotexp ~is valid.
Relaxing $\phi\equiv \phi_{\infty}$ also gives dynamical exponential
couplings in the field equations, but neither these changes nor the
scale factor seem to qualitatively alter our solution discussion
in this $\phat$ regime. The dilaton kinetic term is restored in
$\delta^{\prime}$,
$$
\delta^{\prime} =-{2\over r}\left[ e^{-2\phi} \left[ f^{\prime 2} +
f^2 \ac^2\right] +
{1\over 8} e^{4\phi} {\left( r s^{\prime}\right)^2}
+{1\over 2}{\left( r \phi^{\prime}\right)^2}\right]  ,
\eqn\deqnphib
$$
and similarly in the expressions \meqn -\teqn ~for $m^{\prime}$ and
$T^{\prime}$, while the axion and gauge field equations acquire
additional terms linear in $\phi^{\prime}$:
$$\eqalignno{
& {r^2\over R^2} s^{\prime\prime} +\left[ {\left(2 r+4r^2\phi^{\prime}
\right)\over R^2}+
r^2 e^{\delta}
{d\over dr}\left( { e^{-\delta}\over  R^2}\right)\right] s^{\prime}
- 4 e^{-4\phi}r^2\left[{\pd V\brps \over \pd s} \right]
 &\eqname\seqnpg\cr
&\phantom{
{r^2\over R^2} s^{\prime\prime} +\left[ {\left(2 r+4r^2\phi^{\prime}
\right)\over R^2}+
r^2 e^{\delta}
{d\over dr}\left( { e^{-\delta}\over  R^2}\right)\right] s^{\prime}
}
- 4 e^{-4\phi}\left[ \ec\ww1 -2ff^{\prime} {\ac\over R^2}\right] =0
&\cr
& r^2 \ec^{\prime} + \left( 2r -2r^2\phi^{\prime}\right) \ec -2 f^2
\ac -e^{2\phi}
s^{\prime}\ww1 =0 &\eqname\aeqnpg\cr
& {r^2\over R^2} f^{\prime\prime} +\left[
-{2 r^2\phi^{\prime}\over R^2} +
r^2 e^{\delta}
{d\over dr}\left( { e^{-\delta}\over  R^2}\right)\right] f^{\prime}
+\left[ \ww1 +r^2 {\ac\over R^2}\left( \ac -
e^{2\phi} s^{\prime}\right)\right] f=0 .&\eqname\feqnpg\cr
 }
$$
The equations \seqnpb -\aeqnpb ~and \feqnpb , which were derived from
the axion and gauge field equations, retain their form, and
the inequalities
\sintsign -\seqnpc ~and \flimp ~which followed from them are still
valid if we take $\phi_{\infty}\rightarrow \phi\brr$.

Though the solution scenario explored in connection with
$\phi\equiv\phi_{\infty}$ appears promising when
the scale factor in \dilpotexp ~is convergent, the general
non-Abelian case still requires an analysis of the dilaton equation
$$\eqalign{
{1\over 2}e^{\delta}{d\over dr}\left[ { e^{-\delta}\over  R^2}
r^2 \left( \phat^2\right)^{\prime}\right] =&
 {\left( r\phat^{\prime }\right)^2 \over R^2} +\phat
{\pd V\brps \over \pd\phi} r^2 +\phat\left(
{1\over 2} e^{4\phi} {\left( r s^{\prime}\right)^2\over R^2}\right)\cr
&-2 \phat e^{-2\phi}
\left({f^{\prime 2}\over R^2}
+{\ww1^2\over 2r^2}
-{1\over 2} r^2\ec^2 -{ f^2\ac^2\over R^2}\right) ,\cr
}\eqname\peqnpg
$$
which we have rewritten in a form analogous to \feqnpb.
Since the inequalities derived above
provide no information about the relative magnitudes of
the terms on the right-hand side of \peqnpg, it is difficult
to extract an inequality restricting the behavior of $\phat$, but we
can make some observations. Since $\pd V/\pd \phi < 0$ for $\phat< 0$
and near the minimum, the potential contribution and possibly the
gauge field
contribution to the right-hand side of \peqnpg ~are positive, which
tends to drive $\phat $ away from the vacuum. For $\phat >0$ near the
minimum, these signs reverse and make a monotonic decrease of $\phat $
toward $\phat =0$ more likely. Though our preceeding analysis does not
require such monotonic behavior for the dilaton, it is certainly
consistent with the solution scenario and provides a potentially
viable alternative to the $\phi\equiv \phi_{\infty}$ ansatz.

The complexity of the full string-inspired
theory is well reflected in our inability to proceed any further
analytically. It appears that the question of existence of
non-Abelian solutions with nontrivial dilaton and axion fields
can only be settled by further numerical study. Though we have not
actually obtained numerical solutions in the general case, we have
 considered a strategy for simplifying this four-parameter problem.
In past (two-parameter) shooting problems, we have found the
method of ``shooting to a fitting point''
a convincing way to confirm our
numerical results. In this method, shooting parameters at both
fixed points ($r=0 ;r_h$ and $r=\infty$) are chosen so that the
field equations can be integrated toward a common midpoint, at
which the two trajectories join smoothly if the parameters
correspond to a solution. Though this procedure requires more
than double the original number of shooting parameters, the
deviation of the trajectories at the midpoint can be used to
choose new shooting parameters via a multi-dimensional
Newton-Raphson algorithm. If the initial shooting parameters
are reasonably close
to a solution, this procedure is moderately successful at
converging on the solution, but it is difficult to match the
trajectories with an error comparable to the global tolerance
of the integration routine. Once the neighborhood of a solution
is determined, this appears to be a more promising approach to
finding the solution than the method
employed in the next section, at least for three or more shooting
parameters. We hope to utilize this procedure in the future
to find solutions to the full string-inspired theory.

By closely examining some simplifying ansatzes,
we have been able to narrow the possible solution classes to two:
a massive dilaton coupled to a single magnetic
Yang-Mills degree of freedom,
which we denote EYMD theory, and the gauge field coupled to both
massive dilaton and
massive axion fields. Though we could only speculate about solution
scenarios for the latter theory, in the next section we present
and analyze numerical solutions to the former.

\nobreak\nobreak\nobreak\nobreak\nobreak

{\parindent=0pt\bf\chapter{ EYMD Theory: Regular and Black Hole
Solutions{}~\hfil~~\hfil~~\hfil~~\hfil~~\hfil~~\hfil~~}}

{\parindent=0pt\bf EYMD Equations}

As we observed above, the ansatzes $s^{\prime}\equiv 0$ or
$a^{\prime}\equiv 0$ in the
non-Abelian sector give $a\equiv 0$ and
$s\equiv s_{\infty}$ for both regular and black hole
solutions. To obtain numerical solutions to the resulting EYMD system,
it is instructive to reexpress the dilaton as
$\phi\brr\equiv h\brr /r +\p0$, where
$h/r $, which we also denote
$\phat $, is the deviation of $\phi $ from its
vacuum value. With this change and the
simplifications $a\equiv 0$ and $s=s_{\infty}=b_0 (4\pi n) /3$,
 the remaining gauge field equation becomes
$$
{d\over dr}\left( {f^{\prime}\over RT}\right) +{f\ww1\over r^2}
{R\over T}
-2 {f^{\prime}\left( h/r\right)^{\prime} \over R T} =0,
\eqn\feqnd
$$
and the dilaton equation assumes the form
\def\brpso{\left( \phi,s_{\infty}\right) }

$$
{d\over dr}\left({r^2\bryr^{\prime}\over RT}\right)
-V^{\prime}\brpso r^2 {R\over T}
 +2{\eyr}
\left( {f^{\prime 2}\over R^2}+
{\left[ 1-f^2\right]^2\over 2r^2}\right){R\over T} =0 ,\eqn\heqnd
$$
where $V^{\prime}\equiv \pd V/\pd\phi |_{s=s_{\infty}}$.
The Einstein equations \meqn -\teqn ~for EYMD theory are
$$\eqalignno{
m^{\prime} =\eyr&\left[\Rr2 f^{\prime 2}
+{1\over 2}{\ww1^2\over  r^2}\right] &\eqname\meqnd\cr
+&{1\over 2}\Rr2\left( {h^{\prime}}-\yr\right)^2 +V\brpso r^2 &\cr
r\Rr2{T^{\prime}\over T} = \eyr&\left[-\Rr2\
f^{\prime 2} +
{1\over 2}{\ww1^2\over r^2}\right] -{m\over r} &\cr
\phantom{-}\; -&{1\over 2}\Rr2\left( {h^{\prime}}-\yr\right)^2 +
V\brpso r^2 .
&\eqname\teqnd\cr}
$$
with the auxiliary equation
$$
\delta^{\prime} =-{2\over r}\left[ \eyr f^{\prime 2} +
{1\over 2}\left( {h^{\prime}}-\yr\right)^2\right]  \eqn\deqnd
$$
replacing the $T^{\prime}$ equation for black hole solutions.
To simplify the integration of the equations of motion, we use
the metric derivative relation \deriv ~to express the gauge field
and dilaton equations in a form independent of $T$ or $\delta$:
$$\eqalignno{
&r^2\Rr2 f^{\prime\prime}+
\left[ 2m-\eyr {\ww1^2\over  r}-2 V\brpso r^3
-2r\Rr2\left( {h^{\prime}}-\yr\right)
\right]f^{\prime} &\cr
&\phantom{r^2\Rr2 f^{\prime\prime}}\;
+\ww1 f  =0 &\eqname\feqndd\cr
&r^2\Rr2 h^{\prime\prime}+
\left[ 2m-\eyr {\ww1^2\over r}-2 V\brpso r^3\right]
 \left( {h^{\prime}}-{h\over r}\right)-V^{\prime}\brpso  r^3 &\cr
&\phantom{r^2\Rr2 h^{\prime\prime}}\;
+2{\eyr }\left[ r\Rr2 f^{\prime 2} +{\ww1^2\over 2 r}\right] =0.
&\eqname\heqndd\cr}
$$
 Though it is the $h$-dependent form of the field equations
we choose to integrate, we
pass freely between $\phi $, $\phat $, and $h$  in our analysis
when describing  the dilaton.

{\parindent=0pt\bf Analytical Features and Boundary Conditions}

We can anticipate the general features of the solutions and
boundary conditions from the field equations.The gauge field equation
may be rewritten
$$
{1\over 2}{d\over dr}\left[ {\ep\left( f^2\right)^{\prime}\over RT}
\right] =
\ep\left[{\wp2\over RT}+{R\over r^2 T}\left(f^2-1\right) f^2\right] ,
\eqn\feqnddd$$
Since the right-hand side of this equation is
manifestly positive for $f^2>1$, the only nontrivial
solutions  having finite $f^2$  must satisfy $f^2\le 1$.
 By expanding the left-hand side of the equation,
 we can see
that  solutions satisfy $f f^{\prime\prime} <0$ for $f^{\prime}=0$
and $f^2<1$,
which is characteristic of oscillations about $f=0$. As in past studies
with non-Abelian gauge fields coupled to gravity
(see e.g. [\biz -\bart ]), we expect that solutions will be
classifiable by the number of nodes which occur as  $f$ oscillates.

 {}From the $f$ field equation, we note that
the boundary conditions for the gauge field
include $\left| f\right| =1$ and $f=0$ as $r\rightarrow \infty$.
The asymptotic behavior of the field equations reveal that the latter
condition implies $f\equiv 0$, which for black hole solutions
corresponds to an Abelian
gauge field carrying purely magnetic charge.
Without the dilaton,  this case just reduces to the
Reissner-Nordstrom solution, while with the dilaton we recover the
type of theory studied in [\hohob -\grha].  The
 $f\equiv 0$ case for regular solutions is
 forbidden by boundary conditions at the origin:
regularity of the metric only allows the possibility $\left| f\br0
\right| =1$. If we attempt to set $\left| f\right|\equiv 1$,
corresponding to the theory of a
massive dilaton coupled to Einstein gravity, we find that
both regular and black hole solutions are forbidden:
according to \tcon , $T$ must be monotonically increasing,
but boundary conditions provide the incompatible
restrictions $T\br0 > T\brinf $ and  $T\brh > T\brinf $.
We conclude that for fundamentally non-Abelian solutions
to EYMD theory, we must have
$\left| f\brinf\right|=\left| f\br0\right|=1$ for regular solutions,
$\left| f\brinf\right|=1$ and $f\brh $ unspecified
for black hole solutions,
and $f^2\brr \le 1$ for all solutions.

To better understand the expected behavior of $\phi $, we rewrite
the dilaton equation
$$
{1\over 2}{d\over dr}\left[  { r^2 \left(\phat^2\right)^{\prime}
\over RT}\right]
={\left( r\phat^{\prime}\right)^2\over RT}-\phat
\left[ 2e^{-2\phi } \left(
 {f^{\prime 2}\over R^2} + {\ww1^2\over 2r^2}\right)
-V^{\prime}\brpso  r^2\right] {R\over  T},
\eqn\heqnddd$$
where $\phat \equiv (h/r)$ is the deviation of $\phi$ from the
asymptotic value fixed by the minimum of $V$. For $\phat<0$ near the
the minimum,  $V^{\prime} <0$ and the right-hand side of \heqnddd
{}~is positive-definite, which implies that $\phat^2$ is strictly
increasing. Although $V^{\prime}$ changes sign before it
vanishes as $\phat\rightarrow -\infty$ (cf. fig.~1), the gauge
field contribution to \heqnddd ~is exponentially amplified
in the same limit, so the details of the potential should
not alter the conclusion that the region $\phat <0$ is forbidden.
We can also predict
that solutions exhibit a monotonic decrease from $\phat >0$ to
the vacuum $\phat =0$. This is unambiguous in the $\mu^2=0$ case, for
which \heqnd ~indicates $\phat^{\prime}$ is strictly negative, but
nonzero $V^{\prime}$ introduces the possibility of turning points.
Carrying out the derivatives in
the left-hand side of \heqnddd ~demonstrates that for
$\phat^{\prime} =0$, $\phat^{\prime\prime} <0$ if the gauge field
contribution dominates $V^{\prime}$, while $\phat^{\prime\prime} >0$
if $V^{\prime}$ is dominant. It follows that $\phat^{\prime}=0$
can only occur at a plateau in the former case, but for
$V^{\prime}$ dominant a turning point in $\phat $ is possible. Since
the gauge field contribution is exponentially suppressed and
$V^{\prime}$ grows even larger as $\phat$ increases, it appears that
a turning point is a precursor to diverging $\phat $ and
cannot be a solution feature.  To summarize,
we expect all solutions for which $\phi\brinf $ is finite to be
characterized  by $\phat $ rolling monotonically to zero from above,
and $\left| f\right|$ approaching unity after crossing
at least once through $f=0$.

In addition to providing a sketch of the behavior of the fields for
regular and black hole solutions, the preceeding analysis
 provides insight into the behavior of the fields when we are close
to a solution. To qualify what we mean by ``close'',
we introduce some details of the approach to solving this system
 numerically.  For regular solutions, finite energy density
$(\K2 /\gt )T_{00}$ and regularity of the
 metric at the origin  give the following behavior as
$r\rightarrow 0$:
$$\eqalignno{
  f\brr  &= -1+ {1\over 2}f^{\prime\prime}\br0 r^2 +
{\cal O}\left( r^4\right) &\eqname\fexprg\cr
  h\brr  &= h^{\prime}\br0 r+
{\cal O}\left( r^3\right) &\eqname\hexprg\cr
 2\mr &= {\cal O}\left( r^3\right) &\eqname\mexprg\cr
  \ln T\brr &= {\cal O}\left( r^2\right) ,&\eqname\texprg\cr
}
$$
where $f^{\prime\prime}\br0$ and $h^{\prime}\br0=\phat\br0$ are
parameters we must adjust to ``shoot'' a solution to match the
asymptotic
conditions $\left| f\brinf\right|=1$ and $\phat\brinf =0$, and we
have taken $f\br0 =-1 $ and used the rescaled initial condition
$T\br0 =1$ introduced in  \tdelc. All the terms not shown explicitly
in \fexprg -\texprg ~depend only on the two shooting parameters and
the free parameters $\mu^2$ and $\phi_{\infty}$ associated with $V$.
Similarly, for black hole solutions we can use the metric condition
$1/R^2\brh =0$ for $m\brh = r_{h}/2$ to expand near the horizon:
$$\eqalignno{
  f\brr &= f\brh + f^{\prime}\brh\rrh+{\cal O}\left( r-r_{h}\right)^2 ,
&\eqname\fexpbh\cr
  h\brr &= h\brh + h^{\prime}\brh\rrh+{\cal O}\left( r-r_{h}\right)^2 ,
&\eqname\hexpbh\cr
  m\brr &= r_{h}/2 +m^{\prime}\brh\rrh+{\cal O}\left( r-r_{h}\right)^2
&\eqname\mexpbh\cr
  \dr   &= 0+\delta^{\prime}\brh\rrh +{\cal O}\left( r-r_{h}\right)^2
&\eqname\dexpbh\cr}
$$
where
 $f\brh$ and $h\brh$ are now the shooting parameters, $r_{h}$ is
a free parameter, and the field equations \meqnd -\heqndd
{}~give  the derivatives
$ f^{\prime}\brh, h^{\prime}\brh, m^{\prime}\brh$,  and
$\delta^{\prime}\brh$  as functions of these parameters.

On the basis of past investigations [\biz -\bart ], we expect solutions
(for a particular  choice of free parameters) to exist only
for discrete values of the
shooting parameters. Using the above analysis, we can anticipate
the asymptotic behavior of the fields for a small neighborhood
surrounding those discrete values in shooting
parameter space. Since $f$ exhibits oscillatory behavior for
 $f^2<0$, and $f^2>0$ leads to diverging $f$, we should look
for a parameter range in which $|f|$ approaches  unity at large $r$
and either exhibits a turning point or exceeds $|f|=1$ and diverges.
If for such a parameter range $\phat $ approaches
$0$ and (for $\mu^2$ nonzero)
then undergoes a turning point and diverges,
 or becomes negative and diverges, then our neighborhood should contain
a point  which gives the correct asymptotic behavior.
We explain the shooting procedure in more detail below.

Although the vacuum values $|f\brinf | = 1$ and $\phat\brinf =0 $
are shared by all solutions,
the behavior of the field equations as
$r\rightarrow \infty$ provides  interesting distinctions
between massless and massive dilatons.
For massive dilatons, the leading-order
asymptotic expansions of the fields and metric functions are
\def\hin{h_{\infty}}\def\po{\phi_0}
$$\eqalignno{
 f\brr &\sim \pm\left( -1+ {c\over r} \right) &\eqname\finf\cr
  h\brr &\sim  a e^{-m_{\phi} r} &\eqname\hinf\cr
  \mr &\sim M -{c^2 e^{-2\p0}\over r^3} &\eqname\minf\cr
  \ln T\brr &\sim \ln\left( {1\over T_{0}}\right) +{M\over r}
 &\eqname\tinf\cr
  \dr   &\sim -\delta_{0} + {1\over 2} {c^2e^{-2\p0}\over r^4} ,
&\eqname\dinf\cr}
$$
where $c$ and $a$ are positive constants, $T_0$ and $\delta_{0}$ are
the rescaled metric constants introduced in the discussion of \tdelc,
and the $+(-)$ sign in \finf ~corresponds to an even(odd) number of
nodes in the function $f\brr$.
 Note that the presence of $m_{\phi}$, defined by
$$
m_{\phi}^2\equiv \left[{\pd^2 V\over \pd \phi^2}\right]_{\p0 }
= 2 \alo^2\left({\alo -1}\over \alo+1\right)^2 \mu^4
 , \eqn\mpdef
$$
forces the dilaton
to approach its asymptotic value exponentially as we expect.
For $\alo \not= 1$, we find it convenient to use $m_{\phi}$
in place of $\mu^2$  as a free parameter.
In the massless case $\mu^2 =0$, the expansions are
$$\eqalignno{
 f\brr &\sim \pm\left(-1+ {c\over r} +{1\over 8}
{c\left( 6M+2\hin -c\right)\over r^2}
\right)&\eqname\finfa\cr
  h\brr &\sim  \hin  +{M\hin \over  r} +{1\over 6}
{\hin\left( 8M^2 -\hin^2\right)\over r^2} &\eqname\hinfa\cr
 \mr &\sim M -{1\over 2}{\hin^2\over r}
-{1\over 2}{M \hin^2\over r^2} &\eqname\minfa\cr
  \ln T\brr &\sim \ln\left( {1\over T_{0}}\right) +{M\over r}
+{M^2\over r^2} &\eqname\tinfa\cr
  \dr   &\sim -\delta_{0} + {1\over 2} {\hin^2\over r^2}+
{4\over 3}{M\hin^2\over r^3} ,&\eqname\dinfa\cr}
$$
where we have included the first two nontrivial orders in $1/r$ to
help demonstrate that the constant $\hin $ plays the role of a
Coulombic charge [\dogaa  ] in one metric function
$$
R^2\brr =\left( 1- {2M\over r}+{\hin^2\over r^2} +
\ord\left( 1\over r^3\right) \right)^{-1},\eqn\rinfa
$$
which has the form of the Reissner-Nordstrom solution,
but is absent from the unrescaled form of the other:
$$
{1\over T^2\brr} =\left( 1- {2M\over r}+
\ord\left( 1\over r^3\right) \right) .\eqn\Tinfa
$$
Noting that $\hin =\displaystyle\lim_{r\rightarrow\infty}\left(
-e^{-\delta} r^2\phi^{\prime} /R^2\right)$, we can obtain an
integral expression for $\hin $ [\dogaa ] from \heqnd,
\def\intinf{\int\nolimits_{0}^{\infty} dr}
$$ \hin =\int\nolimits_{r_0}^{\infty} dr
{d\over dr}\left( -e^{-\delta} {r^2\phi^{\prime} \over R^2}\right)
=\int\nolimits_{r_0}^{\infty} dr \>e^{-\delta} \left[
2{\eyr}\left( { f^{\prime 2}\over R^2}+
{\left[ 1-f^2\right]^2\over 2 r^2}\right)
- r^2 V^{\prime} \right].\eqn\hinteqn
$$
We have included $V^{\prime}$ in this expression for the sake of
generality, and have denoted the lower limit $r_0$ to
emphasize that \hinteqn ~applies to both regular and black hole
solutions, since $\left( -e^{-\delta} r^2\phi^{\prime} /R^2\right)$
vanishes at $r_0=r_h$ as well as $r_0=0$. It is difficult to extract
any further information about $\hin$ from this expression alone, but
we can derive a simpler expression for $\hin$ and other useful
relations by
applying the scaling argument techniques of [\hest ] and utilizing
the metric relation \tcon. Before doing so in the next subsection,
 we address an
ambiguity which arises from the definition of $m_{\phi}$.

In the above discussion, we are careful to qualify
``massless dilaton'' by the condition $\mu^2 =V=0$, because
\mpdef ~indicates that there are two ways to achieve massless
dilatons in this theory: $m^2_{\phi}$
vanishes if either $\mu^2 =0$ or $\alo =1$. The latter case,
corresponding to $\p0 ={1\over 2}\ln (3/b_0)\approx 0.830$, is also
special in another respect.
If we consider a power series expansion of $V$ about $\al =\alo$,
we find  that the coefficient of the $n$th order term varies
as $(1/\alo)^{(n-1)}$, so $\alo =1$ marks the radius of convergence of
the potential expansion. This would present a restriction if we chose
to work with the representation $Y\equiv e^{-2\phi}$ for the
 dilaton preferred by some authors, but for
our field choice the power series expansion
 is well-behaved for all $\alo \ge 0$:
$$\eqalign{
V\left(\phat, s_{\infty}\right) =& \mu^4\alo^2\left( {\alo -1\over
\alo +1}
\right)^2\phat^2 +\mu^4\alo^3{\left(\alo -1\right)
\left(\alo -3\right)\over \left(\alo +1\right)}\phat^3 \cr
&+\mu^4\alo^2{\left(  4\left[ 1-11\alo\right] +\alo^2
\left[\alo-3\right]\left[7\alo-33\right]
\right)\over 12\left(\alo +1\right)^2}\phat^4
+\ord\left( \phat^5\right) .\cr}
\eqn\vexp
$$
{}From this expression,  we find
 $$
V\brpso ={1\over 4} \mu^4 \left( \phi-\p0\right)^4
+\ord\left(\left( \phi-\p0\right)^5\right)\eqn\vlim
$$
for $\alo =1$ and $\phi$ near $\p0 ={1\over 2}\ln (3/b_0)$, so
the potential resembles that of a $\lambda \phi^4$ theory with
$\lambda \equiv 3!\mu^4$ in the vicinity of its minimum, while for
all other
positive values of $\alo $ the potential has $\phat^2$ and $\phat^3$
contributions. When we examine the asymptotic behavior of the field
equations for this special case, we find the same leading order
behavior as in \finf -\dinf ~with the exception of the expansion for
$h\brr$,
$$
h\brr\sim -{1\over 2}{c^2e^{-2\p0}\over r^3} -
{1\over 5}{c^2e^{-2\p0}\left(2M+c\right)\over r^4} .
\eqn\hinflam
$$
Note that the behavior of the field is completely determined by the
gauge field to leading order: although the presence of a cubic
$V^{\prime}$ in \heqndd ~eliminates the possibility of nonzero $\hin $
which arises
 in the $\mu^2 =0$ case, it does not otherwise influence the
asymptotic behavior of $h\brr$ until several orders beyond those
shown in
\hinflam. Most significantly,  in the $\alo =1$ case the dilaton field
can only match the boundary conditions if $\phat<0 $ near $r=\infty$,
which from our discussion of \heqnddd ~is
inconsistent with the behavior dictated by the full dilaton equation
for $\phat_0 >0$ or $\phat\brh >0$. The same arguments also apply
to the range $\phat_{max} \le \phat_0 ,\phat\brh \le 0$, where
$\phat_{max}$ is the location of the local maximum of $V$ (see
fig~.1). For the range $\phat < \phat_{max}$, where $V^{\prime}> 0$,
the dilaton field is  not strictly forbidden to increase, but the
exponential amplification of the gauge field term relative to
$V^{\prime}$ in \heqndd ~and \heqnddd ~makes a solution with
$\phat$ increasing toward $\phat=0$ highly unlikely. After checking
this possibility for a wide range of initial conditions,
we believe that no solutions to EYMD theory exist for $\alo =1$.

{\parindent=0pt\bf Further Analysis}

To further explore the EYMD system analytically, we make use of a
scaling technique developed in [\hest ] for regular solutions and
extend it to study black hole solutions. We find that the same results
may be obtained from the metric relation \tcon, but we investigate
the equivalence of the two approaches elsewhere [\cmo ].

The general procedure of [\hest ] involves defining a nonlocal energy
functional \def\ed{e^{-\delta}}
$$
M=\intinf \left( e^{-\delta} m\right)^{\prime} =\intinf
\>\ed\lag^{(0)}\eqn\mfcn
$$
where
$$
\delta\left(\rt\right) =2\int\nolimits_{\rt}^{\infty} dr\left( \sum_{k}
{ U_{k}\brr\over r}\right)
\eqn\dfcn
$$
is a sum of (non-negative) contributions from the kinetic terms $U_k$
of the matter fields. Demanding that $M$ be stationary with respect
to variations of the independent matter fields then yields the
Euler-Lagrange equations when we treat $\lag^{(0)}$ and $\delta$ as
functionals of those fields, and we restrict our attention to
theories for which the action may be written
$$
S\propto\int dt\> dr\left({U\over R^2}+V\right)\ed ,
\eqn\scaleact
$$
where $U$ and $V$ are independent of the metric functions.
The combination $\lag^{(0)}\equiv
m^{\prime}-\delta^{\prime}m$ in \mfcn ~is just $m^{\prime}$
with the metric function $R\brr$ set to unity; with natural units
restored, $\lag^{(0)}$ is equivalent to an effective matter
Lagrangian with $G=0$. By introducing radial scaling transformations
for the independent matter fields $\psi$ of the type
$$
\psi_{\lambda}\brr =\lambda^{n_{\psi}}\psi \left(\lambda r\right) ,
\eqn\pscal
$$
which lead to the decompositions
$$
U_{\lambda}\brr=\sum_k\lambda^{p_k} U_k \left(\lambda r\right) ,
\>\>\>\>\>\>\>\>\lag_{\lambda}^{(0)} =\sum_j
\lambda^{l_j} \lag_{j}^{(0)}\left(\lambda r\right),
\eqn\ulscal
$$
we can find the $\lambda$-dependent energy functional
$M\left(\lambda\right)$ and obtain the constraint
$$\left[{dM \over d\lambda} \right]_{\lambda =1} =
\intinf \sum_j\left[\lag_{j}^{(0)}\left( l_j -1-\sum_k p_k
\delta_k\right)\right]\ed =0 .\eqn\lamcon
$$
Depending on the positivity properties of
$\lag_{j}^{(0)}$  and the various scaling constants $\{ l_j, p_k\}$,
the constraint can be used to establish non-existence theorems or
useful virial-type relationships for the kinetic and potential
contributions of the matter fields. In the present case,
$$\eqalign{
U_f &= \ep\left(f^{\prime}\right)^2 \>\>\>\>\>\>\>\>\>\>
V_f=\ep {\ww1^2\over 2 r^2}\cr
U_{\phi} &= {1\over 2}\left(r\phi^{\prime}\right)^2
\>\>\>\>\>\>\>\>\>\>\>\>\>V_{\phi} =r^2 V\brpso ,\cr}
\eqn\uv
$$
and the constraint $dM /d\lambda |_{\lambda =1} =0$
assumes the form
$$\intinf \left( y^{\prime}_f +
y^{\prime}_{\phi}\right)=
2 \intinf  y^{\prime}_f
-2\intinf \> \delta_f y^{\prime} -2\intinf \ed V_{\phi},\eqn\constr
$$
where $\delta_f$ is the contribution to $\delta$ from $U_f$,
$y\equiv \ed m$, and
$$
y^{\prime}_f\equiv\ed\left( U_f+V_f\right)
\>\>\>\>\>\>\>\>\>\>
y^{\prime}_{\phi}\equiv\ed\left( U_{\phi}+V_{\phi}\right)
\eqn\yfp
$$
are the gauge field and dilaton contributions to $y^{\prime}$.

The approach to black holes in [\hest ] involves defining an effective
Lagrangian $\lag^{(B)}$ on a fixed black hole background and a
different scaling for the matter fields, $\psi_{\lambda}
\left( r/r_h\right)
=\psi (\left( r/r_h\right)^{\lambda} )$. Such an approach yields
a complicated integral relation which generically involves
$\ln ( r/r_h)$ in the integrand. Though still quite useful in
establishing non-existence theorems, the result of the procedure
is not very useful as a pseudovirial relation. Since we are interested
in a means of simplifying integral relations such as \hinteqn, a
relation analogous to \lamcon ~for black holes is more appropriate
for our purposes. To
extend the regular solution analysis, we write the energy
functional as a sum of horizon and non-horizon contributions,
\def\intr0{\int\nolimits_{r_0}^{\infty} dr}
 \def\brro{\left( r_0\right)}
$$
M=\intr0\ed\lag^{(0)} +e^{-\delta\brro} m\brro .
\eqn\mfcnb
$$
By treating $\lag^{(0)}$ , $\delta$ and $\delta\brro$ as functionals
of the matter fields, we again recover the correct equations of motion
from the variation of $M$. The corresponding constraint equation is
$$\eqalign{
\left[{dM \over d\lambda} \right]_{\lambda =1} =
\intr0 \sum_j\left[\lag_{j}^{(0)}\left( l_j -1-\sum_k p_k \delta_k
\right)\right]\ed -e^{-\delta\brro}m^{\prime}\brro r_0 &\phantom{ }\;
\cr-e^{-\delta\brro}m\brro\left(\sum_k p_k \delta_k\brro\right) &=0
.\cr}
\eqn\lamconb
$$
In the notation introduced above, this constraint can be written for
EYMD theory as
$$\eqalign{
\intr0 \left( y^{\prime}_f +
y^{\prime}_{\phi}\right)=&
2 \intr0  y^{\prime}_f
-2\intr0 \> \delta_f y^{\prime} -2\intr0 \ed V_{\phi}\cr
&-2e^{-\delta\brro}m\brro\delta_f\brro -r_0e^{-\delta\brro}
\left( V_f\brro +V_{\phi}\brro\right) ,
\cr}\eqn\constrb
$$
the left-hand side of which is equivalent to
$M-e^{-\delta\brro}m\brro$.

With the addition of the definition $x\equiv -r^2\phi^{\prime}$,
the integrated dilaton equation \hinteqn ~in this notation becomes
$$ \hin =\intr0
{d\over dr}\left[\left( e^{-\delta} -{2y\over r}\right)x\right]
=2\intr0 \left[  y^{\prime}_f -\delta_f y^{\prime} \right]
-\intr0\> r^2\ed V^{\prime}
+2\intinf \> \left(\delta_f y\right)^{\prime}.
\eqn\hinteqnb
$$
Substituting the constraint \constrb ~then gives
$$\eqalign{
\hin =&\intr0 \left(  y^{\prime}_f+y^{\prime}_{\phi}
 \right)
-\intr0 \>r^2\ed \left( V^{\prime}-2V\right) +2
\left[\delta_f y\right]^{\infty}_{0}  \cr
&+r_0e^{-\delta\brro} \left( V_f\brro +V_{\phi}\brro\right)
+2y\brro\delta_f\brro \cr
=&M-e^{-\delta\brro}m\brro -\intr0\> r^2\ed
\left( V^{\prime}-2V\right)\cr
&+r_0e^{-\delta\brro} \left( V_f\brro +V_{\phi}\brro\right) .
\cr}
\eqn\meqh
$$
With $\mu^2=0$, this establishes the simple result $h_{\infty}=M$
for regular solutions  and
\def\eprh{e^{-2\phi\brh}}$$
h_{\infty}=M+r_h e^{-\delta\brh}\left(
\eprh {\wwrh1^{2}\over 2r_{h}^2}
-{1\over 2}\right)
\eqn\hinfbh
$$
for black holes with $r_0=r_h=2m\brh $.

For nonvanishing potential, eq.\meqh ~also gives the interesting
regular solution relation
$$
M=\intinf \> r^2\ed\left( V^{\prime}-2V\right) =
{1\over 4\pi}\int d^3 x\sqrt{ -g_{str}}\ \ep\left(
{\pd V_{str}\over \pd \phi}\right)\eqn\meqvp
$$
where $e^{2\phi} V_{str} \equiv V$ and ``{\it str}''
 describes quantities
in the string frame, which is related to our Einstein frame
by the  conformal transformation ${ g_{\mu\nu}=\ep g^{str}_{\mu\nu}}$ :
$$\eqalign{
\int d^4x&\sqrt{ -g_{str}}\ep\left( R_{str}+4\pd_{\mu}\phi\pd^{\mu}\phi
-4V_{str}\brps +\dots\right) \cr
&{\longmapsto} \int d^4x\sqrt{ -g}\left( R -2\pd_{\mu}\phi\pd^{\mu}\phi
-4V\brps +\dots\right)  .\cr
}
\eqn\strtran
$$
If there is any significance to the suggestive form of \meqvp, it is
not readily apparent.

The regular solution scaling analysis of [\hest ] and its extension to
black hole solutions provide an analytical relation among the solution
parameters and the integrals of the dilaton potential which can be
verified numerically. Since this nontrivial information about the
system must be realized by the field equations, we expect that it
might be obtained directly from them. A closer examination of the
metric relation \tcon ~reveals this to be the case. In the notation
of the scaling discussion, \tcon ~may be rewritten
$$
{d\over dr}\left[{r^2\over R}\left( {1\over T}\right)^{\prime}\right]
=-{d\over dr}\left[{r^2e^{-\delta}\over R^2}\left( \ln T
\right)^{\prime}\right] =2y^{\prime}_f +2 y\delta^{\prime}_f-2
e^{-\delta}V_{\phi} .\eqn\tconb
$$
Using the $T$ equation
$$
{r\over R^2}\left( \ln T\right)^{\prime}=-{1\over R^2}\left(
U_f+U_{\phi}\right)+\left(V_f+V_{\phi}\right)-{m\over r}
\eqn\teqndb
$$
to express the boundary terms, we obtain the relation
$$\eqalign{M+
\left[{r^2e^{-\delta}\over R^2}\left( \ln T\right)^{\prime}
\right]_{r=r_0}=& M-e^{-\delta\brro}m\brro
+r_0e^{-\delta\brro} \left( V_f\brro +V_{\phi}\brro\right)\cr
=&2 \intr0  y^{\prime}_f
+2\intr0 \> \delta^{\prime}_f y -2\intr0 \ed V_{\phi} ,
\cr}\eqn\constrc
$$
which is identical to the scaling result \constrb ~with the
$\delta_f y^{\prime}$ contribution integrated by parts. This
result provides the alternate expressions
$$\eqalignno{h_{\infty} =&M+
\left[{r^2e^{-\delta}\over R^2}\left( \ln T\right)^{\prime}
\right]_{r=r_h} &\eqname\hinfbhb\cr
M=&{1\over 4\pi}\int d^3 x\sqrt{ -g_{str}}\ \ep\left(
{\pd V_{str}\over \pd \phi}\right)
-\left[{r^2e^{-\delta}\over R^2}\left( \ln T\right)^{\prime}
\right]_{r=r_h}
&\eqname\meqvpb\cr}
$$
for eqn.\hinfbh ~and eqn.\meqvp ~generalized to black hole solutions.
Note that the different signs of the kinetic contributions in
$m^{\prime}$ and $r\left( \ln T\right)^{\prime} /R^2$ indicates
that the result from the functional approach \lamconb ~agrees with
\constrc ~only when $r_0$ is the radius of an event horizon, not some
arbitrary radius. It is also interesting to note that the regular
solution parameter values correspond to the $r_h\rightarrow 0$ limit
of the black hole values, which is consistent with solution properties
observed in previous studies of non-Abelian gauge fields coupled to
Einstein gravity [\biz -\kuma ].

We can use the combination of the metric relation \tcon ~and the
dilaton field equation to determine an additional analytical relation.
The combination
$$
{d\over dr}\left[{r^2e^{-\delta}\over R^2}\left(\phi^{\prime}
-\left( \ln T\right)^{\prime}\right)
\right]=r^2 e^{-\delta}\left(V^{\prime}-2V\right)
\eqn\tphicom
$$
can be integrated with the help of \meqh ~to yield a useful
relation between $\phi$ and $T$:\def\brt{\left(\rt\right)}
\def\brhat{\left(\rhat\right)}
$$\eqalign{
\left( \phi\left( r\right) -\phi_{\infty}\right) =& \ln T
\left( r\right) +\left( M-h_{\infty}\right)
\int\nolimits^{r}_{\infty} d\rt {R^2\brt e^{\delta\brt}\over \rt^2} \cr
&+\int\nolimits^{r}_{\infty}  d\rt{R^2\brt e^{\delta\brt}\over \rt^2}
\int\nolimits^{\rt}_{\infty} d\rhat \>\rhat^2 e^{-\delta\brhat}
\left(V^{\prime}-2V\right) .
\cr}
\eqn\tphicomb
$$
In the $\mu^2=0$ case, we have the simple regular solution feature
$\phi\left( r\right) -\phi_{\infty} = \ln T\left( r\right) $,
which can be easily verified
numerically, while for black holes the above equation gives
$$
\left( \phi\left( r\right) -\phi_{\infty}\right) = \ln T
\left( r\right) +\left( M-h_{\infty}\right)
\int\nolimits^{r}_{\infty} d\rt {R^2\brt e^{\delta\brt}\over \rt^2}  .
\eqn\tphicomc
$$
The generic relation for  a nontrivial potential is obtained from
\tphicomb ~by setting $h_{\infty}$ to zero.

{\parindent=0pt\bf Numerical Regular Solutions}

As we observed above, solving the the EYMD equations numerically
is a two-parameter shooting problem. In terms of
the shooting parameters $b\equiv f^{\prime\prime}\br0 /2 $ and
\def\epo{e^{-2\po}}
 $\po\equiv\phi\br0= (h^{\prime}\br0 +\p0 )$
 the boundary conditions \fexprg-\texprg ~become
$$\eqalignno{
  f\left( r\right) =& -1+ br^2 +{1\over 30}b\left( -24b^2\epo -9b +
12V_{0} +4V_{0}^{\prime} \right) r^4 +{\cal O}
\left( r^6\right) &\eqname\fexprga\cr
  h\brr =& \left(\po-\p0\right) r + {1\over 6}
\left( V_{0}^{\prime} -12 b^2\epo \right)r^3 +
{1\over 360}V_{0}\left( 20 V_{0}^{\prime}+3V_{0}^{\prime\prime}\right)
r^5 &\eqname\hexprga\cr
&+{1\over 360}b^2 \epo\left( -576 b^2\epo +288 b - 432V_{0}+
48V_{0}^{\prime}- 36V_{0}^{\prime\prime}\right) r^5
 +{\cal O}\left( r^7\right)  &\cr
  2\mr =& \left(4b^2\epo +{2\over 3}V_{0}\right)r^3
&\eqname\mexprga\cr
&+{1\over 45}\left( b^2 \epo\left[ -144 b +96V_{0}
-48V_{0}^{\prime}\right] +4 V^{\prime 2}_{0} \right)r^5
+{\cal O}\left( r^7\right) &\cr
  \ln T\brr =& -\left(2b^2\epo -{1\over 3} V_{0}\right)r^2
+{1\over 180}\left( 20 V_{0}^{2}+3V_{0}^{\prime 2}\right) r^4
 &\eqname\tzerexpd\cr
 &-{1\over 180}b^2 \epo\left( 288 b^2\epo -144 b
+144 V_{0}+48 V_{0}^{\prime } \right) r^4 +
{\cal O}\left( r^6\right) , &\cr
}
$$
where $V_{0}$, $V_{0}^{\prime}$ and
$V_{0}^{\prime\prime}$ are the potential and its derivatives with
respect to $\phi$ at  $\phi=\phi_{0} $, and the shooting parameters
satisfy $b>0$ and $\po >\p0 $. We evaluate the initial conditions at
$r=10^{-3}$ and use global error
tolerance $10^{-12}$ in an adaptive fifth-order
Runge-Kutta ordinary differential equation solver,
adjusting  $\left( b,\po \right) $ for fixed
$\p0 $ and $m_{\phi}^2$
 and integrating toward $r=\infty $.
For a range of $b$ and $\phi_0$ in the vicinity of a solution, the
fields behave much as we anticipated: the gauge field either undergoes
a turning point at $|f| \lsim 1$ or diverges, and $\phat$ either
undergoes a turning point at $\phat \gsim 0$ or becomes negative and
diverges. In the $\mu^2 =0$ case, the property $\phat^{\prime}\le 0$
excludes $\phat $ turning points, and we instead observe a transition
between monotonic decrease to $\phat\brinf =0$ and
the divergence $\phat\rightarrow -\infty$
over a small parameter range. In terms of $h$, the finite dilaton
behavior is characterized by a turning point after exponential
decrease toward $h\brinf =0$, or the approach toward a positive
constant value $\hin $ when $\mu^2=0$.

In the massive case, finding a neighborhood of some discrete point
$\left( b,\po \right) $ which exhibits these properties does not
guarantee that one has found a legitimate solution to the EYMD system.
The key to determining whether the results of the shooting procedure
constitute valid solutions lies in the
 exponential behavior of $h$.
For $h$ to decay exponentially to zero, the gauge field
coupling term in \heqndd ~must be insignificant. Since this
contribution is positive definite at finite radius, we must rely on its
algebraic approach to zero via \finf ~to satisfy this condition.
As $\mu^2$ is increased from zero, the radius at which $h$
exponentially decays behaves roughly as $r\sim 1/ m_{\phi}$, and
eventually encroaches upon the fixed region where the gauge
contribution algebraically decays.  In other words, the screening of
the Coulombic dilaton charge $\hin $
 occurs at a radius $r\sim 1/m_{\phi}$ which approaches the zone where
the local magnetic charge density vanishes. According to \heqndd,
once $m_{\phi}$ is large enough that these regions overlap, the
gauge field source drives the dilaton away from its vacuum value
and solutions are not possible. For a range of $m_{\phi}$ near this
overlap, $h$ appears to decay exponentially and satisfy
the numerical solution criteria, but close examination reveals
deviations from exponential behavior which prohibit extrapolation to
$h\brinf $. To determine a maximum allowable
mass $(m_{\phi})_{max}$ in practice, we must set a
limit on the
deviation of $h$ from the behavior required to match the boundary
conditions. By considering next-to-leading order terms in \heqndd, we
find that
$$
h\sim a r^{-m_{\phi} M} e^{-m_{\phi} r}\eqn\hinfexp
$$
describes the asymptotic behavior of the dilaton more precisely than
\hinf, so that
$$
{\delta h^{\prime\prime}\over h^{\prime\prime}}
\equiv {1\over h^{\prime\prime}} \left[ h^{\prime\prime} - m_{\phi}^2
\left( 1 + {2 M\over r}\right) h\right]
\eqn\dlhpa
$$
gives a fair measure of the deviation from ideal behavior. Another
useful quantity is
$$
{\Delta h^{\prime\prime}\over h^{\prime\prime}}
\equiv -{1\over h^{\prime\prime}}\left({2 R^2\over r}{\eyr }\left[
 {f^{\prime 2}\over R^2} +{\ww1^2\over 2 r^2}\right] \right) ,
\eqn\dlhpb
$$
which directly measures the contribution of the gauge field coupling
term to $h^{\prime\prime}$. In the asymptotic regime of a valid
solution, we expect
$|\Delta h^{\prime\prime}/ h^{\prime\prime} | \ll h^{\prime\prime}
\ll 1$ and  \dlhpb ~to be comparable to \dlhpa, but the
maximum acceptable value of either quantity is somewhat ambiguous.
Since the size of the contributions to the dilaton equation which are
not accounted for by the deviation formula
\dlhpa ~are roughly of order $h^{\prime\prime} /r^2$, and
${\overline r}\sim 10^3$ is the characteristic radius at which $f$
obeys the asymptotic form \finf, we adopt the
criterion $|\delta h^{\prime\prime}/ h^{\prime\prime}|
\lsim 1/{\overline r}^2 \approx 10^{-6}$. We find that this
criterion gives the consistent result $(m_{\phi})_{\max}
\sim 1/{\overline r}\sim 10^{-3}$
 for the classes of solutions we investigate.

When actually obtaining solutions, we adjust the the shooting
parameters until the solution bracketing conditions (the turning
points and divergent behavior of $f$ and $h$
which characterize the solution neighborhood) indicate that the
intervals containing the discrete solution values are smaller than
our machine accuracy. To achieve this precision as $m_{\phi}$ is
increased, we truncate $h$ by taking $h\rightarrow 0$ or by
attaching an exponential tail $h\rightarrow a e^{-m_{\phi} r}$ at
the final turning point;
this allows the integration to proceed so that the $f$ bracketing
condition can be determined. To justify this procedure, we apply the
deviation criterion \dlhpa ~at the turning point and verify that $h$
 behaves according to \hinfexp ~to better than one part in $10^6$.
Since $h$ tends algebraically to $ h_{\infty}$ rather than
experiencing a turning point when the dilaton is massless, the
truncation procedure is unnecessary and the numerical pitfalls posed
 by the final term in \heqndd ~disappear.

The results of the shooting procedure for the choice $\p0 =0$ are
displayed in fig.~2, with the massless and $(m_{\phi})_{max}$
solution properties summarized in table~1. Like the results of
previous studies [\biz -\bart ], solutions can be classified by the
number of nodes $k$ exhibited by the non-Abelian gauge field function
$f$. Though an infinite number of solution classes exist, we focus our
attention on the lowest odd- and even-$k$ classes. For both of these
classes, we performed an identical shooting procedure with the
toy potential $V\left(\phi\right) = m_{\phi}^2\phi^2 /2$ for
comparison. As table~1 indicates, the results agree to better than
one part in $10^4$ for the narrow range of allowable dilaton masses,
which indicates that the higher-order $\phat $ terms in the potential
expansion \vexp ~are negligible for this choice of $\p0$.
Though the total mass $M$ measurably increases as $k$ increases and
$m_{\phi}$ varies over the allowed range, the only substantial change
in the function plots occurs for $h\brr$.  It exhibits a maximum at
small radius and
approaches $h_{\infty}$ at large $r$, where the Coulombic dilaton
charge is exponentially screened progressively closer to
the decay zone of the gauge field. The plot of the
actual dilaton field $h\brr /r$, along with the gauge function $f\brr$
and the mass-energy $m\brr$, exhibits nontrivial variation only in
the decades surrounding $r=1$: the characteristic radius of the
solution is fixed by the string coupling, which in our dimensionless
variables mimics the choice $\gt \equiv 2/\sqrt{\alpha^{\prime}} =1$.
Note that the maximum dilaton mass, which we only determine to
one decimal place using our imprecise criterion, decreases as the
characteristic radius of the gauge field decay zone increases with
increasing $k$.

\def\tablerule{\noalign{\hrule}}
\vskip .25in
\vbox{\tabskip=0pt \offinterlineskip
\halign to473pt{\strut#& \vrule#\tabskip=1em plus2em
&\hfil#& \vrule#
&\hfil#\hfil& \vrule#
&\hfil#\hfil& \vrule#
&\hfil#\hfil& \vrule#
&\hfil#\hfil& \vrule#
&\hfil#\hfil& \vrule#
&\hfil#\hfil& \vrule#\tabskip=0pt\cr\tablerule
&&\multispan3 &\multispan9\hfil
EYMD Regular Solutions $\phi_{\infty}=0$
\hfil &\multispan1 &\cr\tablerule
&&\omit\hidewidth $k$\hidewidth&&
\omit\hidewidth $m_{\phi}$ \hidewidth&&
\omit\hidewidth $b$\hidewidth&&
\omit\hidewidth $\phi_0$\hidewidth&&
\omit\hidewidth $M$\hidewidth&&
\omit\hidewidth $h_{\infty}$\hidewidth&&
\omit\hidewidth $\ln T\br0$\hidewidth &\cr\tablerule
&&1&&0&&1.0755243&&0.9322839&&0.5769856&&0.5769856&&0.9322839&\cr
&&1&&$3\times 10^{-3}$&&1.0716772&&0.9304842&&0.5775049&&0&
&0.9322771&\cr
&&1\rlap*&&$3\times 10^{-3}$&&1.0717784&&0.9305347&&0.5774929&&0&
&0.9322762&\cr\tablerule
&&2&&0&&8.3620815&&1.7927935&&0.6848332&&0.6848332&&1.7927935&\cr
&&2&&$2\times 10^{-3}$&&8.3391611&&1.7914176&&0.6853146&&0&
&1.7928245&\cr
&&2\rlap*&&$2\times 10^{-3}$&&8.3395171&&1.7914400&&0.6853065&&0&&
1.7928145&\cr\tablerule
\noalign{\smallskip}
&\multispan9* (run with $V\left(\phi\right) = {1\over 2}m_{\phi}^2
\phi^2$ for comparison)\hfil\cr}}
\vskip .075in
\centerline{ Table 1.}
\vskip .15in

The solution parameters $M$, $h_{\infty}$ and $\ln T\br0$ in table~1
are determined with the aid of the asymptotic expansions \finf -\tinf
{}~and \finfa -\tinfa. In the massless case, they provide a good check of
some of the analytical relations determined above. In particular,
$h_{\infty} =M$ to better than seven figures in accordance with
\hinfbh, while the prediction of \tphicomc ~that
$\phat\brr=\ln T\brr$ is confirmed at $r=0$ (and other points) to
an accuracy exceeding the global error tolerance $10^{-12}$. The
agreement of numerical and analytical results is
compelling evidence for the general accuracy of our shooting method;
the numerical results of other authors who recently considered
the massless case [\dogaa ] do not clearly exhibit these relations,
though they are in general agreement with table~1.

As values of $\p0 $ in the range $-0.8 \lsim \p0\lsim 0.8$ are used
in our procedure, solutions
appear to be related to the $\p0 =0$ solutions by the scaling
of the radius and some physical parameters. To understand and quantify
the scaling, we consider the simplified case of the massless dilaton.
When we ignore the axion and the dilaton-axion potential in the
action \actc, the dilaton explicitly appears only in the exponential
coupling to the non-Abelian field strength. If we absorb the constant
$e^{-2\p0}$ into the gauge coupling $g^2$ and then rewrite the theory
in terms of dimensionless variables and parameters, we recover the
$\p0 =0$ theory but with $\gt\rightarrow e^{-\p0}\gt$. {}From the
definitions of the dimensionless quantities \dimquant -\mhatdef, we
therefore expect the
 radial structure of the $\p0\not= 0$ solutions to scale
according to $r\rightarrow e^{-\p0}r$. The mass-energy should similarly
scale as $m\rightarrow e^{-\p0}m$, but the amplitude of the dilaton
deviation field $\phat$ and the gauge field should remain unchanged.
It follows that the dilaton shooting parameter shifts such that
$\phat\br0 =\po-\p0$ is unaffected, while the gauge field parameter
scales as $b\rightarrow be^{2\p0}$ to compensate for the scaling of the
intital value of $r^2$. Though the introduction of the potential and
its complicated dependence on $\p0$ changes this picture, solutions
approximately obey the same scaling relations for the range of $\p0$
examined, with $(m_{\phi})_{max}\rightarrow e^{\p0}(m_{\phi})_{max}$
as expected from \dimquant ~and the relationship $(m_{\phi})_{max}
\sim 1/{\overline r}$ discussed above. As $\p0$ approaches the
critical value ${1\over 2}\ln (3/b_0) \approx 0.830$, at which the
potential
 \vexp ~reduces to the $\lambda \phat^4$ form \vlim ~to leading order,
solutions become harder to obtain numerically and appear to be
forbidden at ${1\over 2}\ln (3/b_0)$ for reasons examined above.
For $\p0 $ above this critical value, the leading order
$m_{\phi}^2\phat^2$  term is restored in the potential expansion and
solutions are again possible, though we do not explore the scaling
properties of this solution region in depth.

{\parindent=0pt\bf Numerical Black Hole Solutions}

To find numerical black hole solutions, we use the conditions
$$\eqalignno{
  f^{\prime}\brh &= { -\wwrh1 f\brh \over
\left(  r_{h}-\eprh{\wwrh1^2 /  r_{h}}
-2V\left(\phi\brh\right) r_{h}^3 \right) }
&\eqname\frh\cr
  h^{\prime}\brh &=
{V^{\prime}\left(\phi\brh\right) r_{h}^3  -\eprh {\wwrh1^2 / r_h}
\over
 \left(  r_{h}-\eprh{\wwrh1^2 /  r_{h}}
-2V\left(\phi\brh\right) r_{h}^3 \right) } +{h\brh\over r_h}
&\eqname\hrh\cr
   m^{\prime}\brh &= \eprh {\wwrh1^{2}\over 2r_{h}^2}
 +V\left(\phi\brh \right) r_{h}^2&\eqname\mrh\cr
 \delta^{\prime}\brh &= -{2\over r_h}
 \left(  f^{\prime 2}\brh +{1\over 2}
\left[ h^{\prime}\brh -{h\brh\over r_h}\right]^2 \right)
&\eqname\drh\cr}
$$
on the horizon, and use $f\brh$ and $h\brh= r_h
\left(\phi\brh -\p0\right) $ as shooting parameters for
$r_{h}=1$. The asymptotic properties of the fields, which we use to
locate the neighborhood of a solution in shooting parameter
space, are identical to the regular solution properties, so we follow
 precisely the same shooting procedure detailed above.
In particular, the same truncation of $h$ and
maximum dilaton mass criterion are used to determine solutions in
the massive case.

\vskip .25in
\vbox{\tabskip=0pt \offinterlineskip
\halign to473pt{\strut#& \vrule#\tabskip=1em plus2em
&\hfil#& \vrule#
&\hfil#\hfil& \vrule#
&\hfil#\hfil& \vrule#
&\hfil#\hfil& \vrule#
&\hfil#\hfil& \vrule#
&\hfil#\hfil& \vrule#
&\hfil#\hfil& \vrule#\tabskip=0pt\cr\tablerule
&&\multispan3 &\multispan9\hfil
EYMD Black Hole Solutions $\phi_{\infty}=0$
\hfil &\multispan1&\cr\tablerule
&&\omit\hidewidth $k$\hidewidth&&
\omit\hidewidth $m_{\phi}$ \hidewidth&&
\omit\hidewidth $f\brh$\hidewidth&&
\omit\hidewidth $\phi\brh$\hidewidth&&
\omit\hidewidth $M$\hidewidth&&
\omit\hidewidth $h_{\infty}$\hidewidth&&
\omit\hidewidth $\delta\brh$\hidewidth &\cr\tablerule
&&1&&0&&$-.5935482$&&0.4422713&&0.8367063&&0.5121333&&0.2418818&\cr
&&1&&$2\times 10^{-3}$&&$-.5939426$&&0.4416375&&0.8369759&&0&
&0.2423254&\cr
&&1\rlap*&&$2\times 10^{-3}$&&$-.5939383$&&0.4416507&&0.8369721&&0&&
0.2423185&\cr\tablerule
&&2&&0&&$-.1320851$&&0.5445464&&0.8650727&&0.5748328&&0.1510324&\cr
&&2&&$3\times 10^{-4}$&&$-.1321129$&&0.5444857&&0.8651225&&0&
&0.1510935&\cr
&&2\rlap*&&$3\times 10^{-4}$&&$-.1321128$&&0.5444860&&0.8651223&&0&&
0.1510932&\cr\tablerule
\noalign{\smallskip}
&\multispan9* (run with $V\left(\phi\right) = {1\over 2}m_{\phi}^2
\phi^2$ for comparison)\hfil\cr}}
\vskip .075in
\centerline{ Table 2.}
\vskip .15in

We again examine only the $k=1$- and $k=2$-node solution classes
for the choice $\p0 =0$ and compare results for the full potential
\dilpot ~and the toy potential $V\left(\phi\right) = m_{\phi}^2
\phi^2 /2$; the results are shown in fig.~3 and table~2. Again only
the function $h$ varies significantly as $m_{\phi}$ increases over
the small allowed interval, with $h$ approaching $h_{\infty}$ and
then vanishing exponentially at progressively smaller radius. The
solution structure is nearly identical to the $r\ge 1$ portion of
the regular solutions, which
in part reflects the occurrence of the event horizon at the
characteristic radius $r=1$ of the dimensionless system. A closer
 examination of
the regular solution functions reveals a sharp peak in the metric
function $R\brr$ near $r=1$, where $2m\brr /r$ closely approaches
unity, thus demonstrating that even the regular solutions are
strongly gravitating. As we might expect from previous work with
black hole solutions to theories with non-Abelian gauge fields
[\biz-\kuma ],
the black hole solutions reduce to the regular solutions for the same
choice of $\{ \p0, m_{\phi}\}$ in the limit $r_h\rightarrow 0$.

Though the total mass-energy and dilaton charge do not obey the
simple relationship $h_{\infty}= M$ enjoyed by the massless dilaton
regular solutions, the analytical result \hinfbh ~does relate
$h_{\infty}$ to $M$, $\delta_0=\delta\brh$ and the integration
parameters $r_h$, $f\brh$ and $\phat\brh $. Again, the results of our
 numerical procedure verify a nontrivial relationship between the
physical parameters of solutions to better than seven significant
figures. Black hole solutions for the massive dilaton
also exhibit the approximate scaling properties explored above for
non-zero $\p0 $, but the specific relations for $r$, $m\brr$ and
$(m_{\phi})_{max}$ only hold when we scale the horizon radius
according to $r_h\rightarrow e^{-\p0}r_h$.

\nobreak

{\parindent=0pt\bf\chapter{Conclusions~~\hfil~~}}

In this paper we have studied static, spherically symmetric regular
and black hole solutions to $SU(2)$ gauge theory coupled to a
massive dilaton, massive axion, and Einstein gravity. Our intentions
 have been two-fold: to explore solutions in the physically relevant
 context of low-energy string theory with massive scalar fields, and
to determine whether ``stringy'' scalar fields lead to non-Abelian
solutions with primary hair and good prospects for stability. After
analyzing all the possibilities for fundamentally non-Abelian
solutions, we found strong
numerical evidence for regular and black hole solutions of a massive
dilaton coupled to the Yang-Mills field (EYMD
 theory), and established a deeper understanding of certain
solution existence techniques [\hest ] in the course of exploring the
solutions analytically.
Though the case of a massive axion coupled to the gauge field appeared
promising, we found that the full theory, which describes
a massive dilaton and massive axion coupled to a dyonic non-Abelian
configuration,
is the only other situation which can admit solutions. We presented
no numerical evidence for such solutions, but we were able to
construct a consistent solution scenario.

An important issue that we have not addressed in depth is the
stability of our solutions. As we noted above, the primary hair
solutions to EYMD theory are
 structurally very similar to the solutions of
EYM [\bart, \biz-\kuma ] and EYMH [\grmaon ] theories, which
have been interpreted as generalized sphalerons and are generically
unstable.  Though examples of stable solutions with non-Abelian
structure
 have been found, including ``black holes inside magnetic monopoles''
[\lenawe -\brfoma ] and Skyrmion black holes [\hedrstb -\bich ],
such solutions typically possess a net gauge or topological charge
(and the resultant imprint at spatial
infinity) which is lacking in our solutions. These observations, and
the linear analysis of [\lamaa ] which established the instability of
EYMD solutions for a massless dilaton, make the stability of our
massive dilaton solutions very unlikely.

In light of this conclusion, one might question the relevance of
pursuing numerical solutions to the full theory. As the only examples
 of gravitating $SU(2)$ solutions with both magnetic and electric
fields, such solutions would be interesting in
their own right, but the lack of a net electric charge (which follows
here from the asymptotic behavior of the field equations) would appear
{\it not} to improve the chances of stability. It is conceivable,
 however, that the structure arising from the coupled electric and
magnetic charge densities substantially modifies the sphaleron
character of EYMD solutions, even in the absence of a net charge.
Only in such circumstances, it seems, could we reasonably hope for
stable solutions. Since it requires a four-parameter shooting
procedure, the task of {\it finding} such solutions could present
enough obstacles that these questions might remain unanswered.
Though we have not yet attempted to obtain such solutions, we
have described a strategy which might simplify this formidable
 task. We hope to test the efficacy of this strategy in future
investigations of string-inspired non-Abelian dyonic solutions.

\FIG\fa{ The dilaton-axion potential \dilpot, which is of the form
 used in the study of SUSY-breaking via gaugino condensation in
string theory. In (a), the rescaled axion field ${\wdhat s}
\equiv s(3/4\pi b_0)$ is fixed at one of the degenerate minima
${\wdhat s} ={\wdhat s}_{\infty}=n$ for integer $n$, and the
 dilaton at spatial infinity is chosen to be $\phi_{\infty} =0$.
 Solutions to
Einstein-Yang-Mills-Dilaton theory correspond to $\phi$ rolling
monotonically to the minimum from the right, confined to a region
where $V$ is well approximated by leading-order $\left(\phi
-\phi_{\infty}\right)$ behavior. In (b), the dilaton is fixed at
$\phi=\phi_{\infty}$ and the potential assumes the form
$V\propto 2\left( 1-\cos\left( 2\pi{\wdhat s}\right)\right)$.
A possible solution scenario for the full theory involves the
axion traversing one of the maxima in the ${\wdhat s}$ direction
and monotonically approaching an adjacent minimum as
$\phi\rightarrow\phi_{\infty}$ from above. The fact that such
non-Abelian dyonic solutions are conceivable critically depends on
the presence of both the massive axion and massive dilaton fields.}

\FIG\fb{ One- and two-node regular solutions to
Einstein-Yang-Mills-Dilaton theory for the dilaton potential of
fig.~1a. The connection function $(1+f\brr )$, total mass-energy
$m\brr$,
dilaton $\phi\brr$ and $h\brr =r\phi\brr$ are plotted as functions of
 radius for a range of dilaton masses $0\le m_{\phi} <\left( m_{\phi}
\right)_{max}$. The maximum mass $\left( m_{\phi}\right)_{max}$
is reached as the exponential screening of $h\brr$, which
asymptotically resembles a Coulombic charge
$h_{\infty}\equiv h\brinf$ in the $m_{\phi}=0$ case,
occurs near the zone where the local magnetic gauge charge vanishes.
The other functions do not vary appreciably in this mass range.
The exact correspondence between $h_{\infty}$ and $M\equiv m\brinf$,
as well as $\phi\brr-\phi_{\infty}$ and the metric function
$\ln T\brr$, is proven analytically in Section 5 for the
massless dilaton case.}

\FIG\fc{ One- and two-node black hole solutions to
Einstein-Yang-Mills-Dilaton theory for horizon radius $r_h =1$ and
the dilaton potential of fig.~1a. As in the regular case, only
the dilaton function $h\brr =r\phi\brr $ varies appreciably as
the dilaton mass
approaches a maximum value. Though $h_{\infty}$ and $M$ are not
 equal for $m_{\phi} =0$, an analytic expression is derived in
Section 5 which relates these quantities to other global
solution parameters. }

\ack
I would like to thank Brian Greene for his guidance
and for suggesting the problem, and Alfred Shapere for
several helpful discussions.
 While this paper was being completed, I received a
preprint by  E.E.~Donets and D.V.~Gal'tsov,
``Stringy Sphalerons and Non-Abelian Black Holes'' [\dogaa ],
which overlaps with some of this work, and I recently became aware of
related papers by G.~Lavrelashvili and D.~Maison [\lamaa-\lamab ]
and P.~Bizon [\bizob ]. This work was supported in part by the
National Science Foundation.
\vfil
\refout\figout
\end